\documentclass[aps,prd,twocolumn,showpacs,superscriptaddress,groupedaddress]{revtex4}
\pdfoutput=1

\usepackage{preprintcover}
\PreprintCoverPaperTitle{Search for squarks and gluinos with the ATLAS detector in final states with jets and missing transverse momentum using 4.7 \ifb~of  $\sqrt{s}=7\TeV$ proton-proton collision data}
\PreprintIdNumber{CERN-PH-EP-2012-195}
\PreprintCoverAbstract{
A search for squarks and gluinos in final states containing jets, missing
transverse momentum and no high-\pt~electrons or muons is presented. 
The data represent the complete sample recorded in 2011 by the ATLAS experiment in $7\TeV$ proton-proton collisions at the Large Hadron Collider, with a total integrated luminosity of $4.7~\mathrm{fb}^{-1}$.
No excess above the Standard Model background expectation is observed.
Gluino masses below 860~GeV and squark masses below 1320~GeV are excluded at the 95\% confidence level in simplified models containing only squarks of the first two generations, a gluino octet and a massless neutralino, for squark or gluino masses below 2~TeV, respectively.
Squarks and gluinos with equal masses below 1410~GeV are excluded.
In MSUGRA/CMSSM models with $\tan\beta=10$, $A_0=0$ and $\mu> 0$, squarks and gluinos of equal mass are excluded for masses below 1360~GeV. Constraints are also placed on the parameter space of SUSY models with compressed spectra.
These limits considerably extend the region of supersymmetric parameter space excluded by previous measurements with the ATLAS detector.
}
\PreprintJournalName{Physical Review D}

\usepackage{graphicx}
\usepackage{epstopdf}
\usepackage{atlasphysics}
\usepackage{subfigure}
\usepackage{multirow}
\usepackage{rotating}
\usepackage{endnotes}

\usepackage[colorlinks,breaklinks,pdftitle={ATLAS preprint},pdfauthor={The ATLAS Collaboration}]{hyperref}  
\hypersetup{linkcolor=blue,citecolor=blue,filecolor=black,urlcolor=blue}

\newcommand{\meff}{\ensuremath{m_{\mathrm{eff}}}}

\makeatletter
\renewcommand{\p@subfigure}{\arabic{figure}}
\renewcommand{\thesubfigure}{\alph{subfigure}}
\renewcommand{\@thesubfigure}{(\thesubfigure)}
\makeatother

\newcommand{\anactualptmissvecwithapinit}{{\vec{P}}_\mathrm{T}^\mathrm{\; miss}}

\newcommand{\ourvecptmiss}{\ensuremath{\anactualptmissvecwithapinit}}
\newcommand{\ourdeltaphishort}{\Delta\phi}
\newcommand{\ourdeltaphifull}{\ourdeltaphishort(\textrm{jet$_i$},\ourvecptmiss)_\mathrm{min}}

\newcommand{\alpgen}{{\tt ALPGEN}}

\newcommand{\herwig}{{\tt Herwig}}
\newcommand{\pythia}{{\tt PYTHIA6}}

\newcommand{\isajet}{{\tt ISAJET}}

\newcommand{\isasusy}{{\tt ISASUSY}}
\newcommand{\jimmy}{{\tt JIMMY}}
\newcommand{\sherpa}{{\tt SHERPA}}

\newcommand{\cteq}{{\tt CTEQ6.6}}
\newcommand{\mstw}{{\tt MSTW2008}}

\newcommand{\ourintlumi}{{$4.7~\ifb$}}
\newcommand{\ourpt}{p_\mathrm{T}}

\newlength{\figwidth}
\title{Search for squarks and gluinos with the ATLAS detector in final states with jets and missing transverse momentum using 4.7 \ifb~of  $\sqrt{s}=7\TeV$ proton-proton collision data}

\begin{document}
\author{The ATLAS Collaboration}
\title{Search for squarks and gluinos with the ATLAS detector in final states with jets and missing transverse momentum using 4.7 \ifb~of  $\sqrt{s}=7\TeV$ proton-proton collision data}
\begin{abstract}A search for squarks and gluinos in final states containing jets, missing
transverse momentum and no high-\pt~electrons or muons is presented. 
The data represent the complete sample recorded in 2011 by the ATLAS experiment in $7\TeV$ proton-proton collisions at the Large Hadron Collider, with a total integrated luminosity of $4.7~\mathrm{fb}^{-1}$.
No excess above the Standard Model background expectation is observed.
Gluino masses below 860~GeV and squark masses below 1320~GeV are excluded at the 95\% confidence level in simplified models containing only squarks of the first two generations, a gluino octet and a massless neutralino, for squark or gluino masses below 2~TeV, respectively.
Squarks and gluinos with equal masses below 1410~GeV are excluded.
In MSUGRA/CMSSM models with $\tan\beta=10$, $A_0=0$ and $\mu> 0$, squarks and gluinos of equal mass are excluded for masses below 1360~GeV. Constraints are also placed on the parameter space of SUSY models with compressed spectra.
These limits considerably extend the region of supersymmetric parameter space excluded by previous measurements with the ATLAS detector.
\end{abstract}
\maketitle

\section{Introduction}
\label{sec:intro}
Many extensions of the Standard Model (SM) include heavy colored
particles, some of which could be accessible at the Large Hadron Collider (LHC) \cite{LHC:2008}.  The squarks and
gluinos of supersymmetric (SUSY) theories~\cite{Miyazawa:1966,Ramond:1971gb,Golfand:1971iw,Neveu:1971rx,Neveu:1971iv,Gervais:1971ji,Volkov:1973ix,Wess:1973kz,Wess:1974tw} form one class of such
particles.  This paper presents a new ATLAS search for squarks and gluinos in final states containing only jets and large missing transverse momentum.
Interest in this final state is motivated by the large number of $R$-parity conserving
models, including MSUGRA/CMSSM scenarios~\cite{Fayet:1976et,Fayet:1977yc,Farrar:1978xj,Fayet:1979sa,Dimopoulos:1981zb}, in which squarks, $\squark$, and gluinos, $\gluino$, can be produced in pairs \{$\gluino\gluino$, $\squark \squark$, $\squark\gluino$\}) and can generate the final state of interest through their direct (
$\squark \rightarrow q\ninoone$ and $\gluino\rightarrow q \bar{q} \ninoone$) and cascade decays to weakly interacting neutralinos, $\ninoone$, which escape the detector
unseen.
`Squark' here refers only to the superpartners of the four light-flavour quarks.
The analysis presented here is based on a study of final states which are reconstructed as purely hadronic. Events with reconstructed electrons or muons are
vetoed to avoid overlap with a related ATLAS search \cite{ATLAS:2011ad} that requires them. 
The term `leptons' is therefore used in this paper to refer only to reconstructed electrons and muons, and does not include $\tau$ leptons.
Compared to previous studies~\cite{Aad:2011ib},
this updated analysis uses the full dataset (\ourintlumi{}) recorded at $\sqrt{s}=7\TeV$  in 2011 and extends the sensitivity of the search by selecting final state topologies with higher jet multiplicities. 
The search strategy is optimized for
maximum discovery reach in the $(m_{\gluino},m_{\squark})$-plane (where $m_{\gluino},m_{\squark}$ are the gluino and squark masses, respectively) for a range
of models. This includes a simplified model in which all other supersymmetric particles, except for
the lightest neutralino, are given masses beyond the reach of the LHC.
Although interpreted in terms of SUSY models, the main results of this analysis (the data and expected background event counts in the signal regions) are relevant for constraining any model of new physics that predicts the production of jets in association with missing transverse momentum.  

The paper begins with a brief description of the ATLAS detector (Section~\ref{sec:ATLAS}), followed by an overview of the analysis strategy (Section~\ref{sec:strategy}). This is followed by short descriptions of the data and Monte Carlo (MC) simulation samples used (Section~\ref{sec:samples}) and of the trigger strategy (Section~\ref{sec:trigger}). Section~\ref{sec:objects} describes the physics object definitions.  Section~\ref{sec:evclean} describes the event cleaning techniques used to reject non-collision backgrounds, while Section~\ref{sec:evsel} describes the final event selections and resulting event counts. Section~\ref{sec:back} describes the techniques used to estimate the SM backgrounds, with the systematic uncertainties summarized in Section~\ref{sec:systematics}. Section~\ref{sec:stats} describes the statistical model used to interpret the observations and presents the results in terms of constraints on SUSY model parameter space. Finally Section~\ref{sec:conc} summarizes the main results and conclusions.

\section{The ATLAS Detector}
\label{sec:ATLAS}

The ATLAS detector~\cite{Aad:2008zzm} is a multipurpose particle
physics apparatus with a forward-backward symmetric cylindrical
geometry and nearly 4$\pi$ coverage in solid angle~\cite{atlascoordinates}.
 The layout of the detector features four superconducting magnet systems, which comprise a
thin solenoid surrounding inner tracking detectors and three large
toroids used in a large muon spectrometer.
Located between these two detector systems, the calorimeters are of particular importance to this analysis.  In the pseudorapidity
region $\left|\eta\right| < 3.2$, high-granularity liquid-argon (LAr)
electromagnetic (EM) sampling calorimeters are used.  An iron/scintillator-tile
calorimeter provides hadronic coverage over
$\left|\eta\right| < 1.7$.  The end-cap and forward regions,
$1.5 < \left|\eta\right| < 4.9$, are instrumented with LAr calorimeters
for both EM and hadronic measurements.

\section{Analysis Strategy}
\label{sec:strategy}

This analysis aims to search for the production of heavy SUSY particles decaying into jets and neutralinos, with the latter creating missing transverse momentum (\met{}). Because of the high mass scale expected for the SUSY signal, the `effective mass', $\meff$ (defined below), is a powerful discriminant between the signal and most SM backgrounds. 
The requirements used to select jets and leptons (which are referred to as physics objects) are chosen to give sensitivity to a broad range of SUSY models. 
In order to achieve maximal reach over the
$(m_{\gluino},m_{\squark})$-plane, six analysis channels are defined. Squarks typically generate at least one jet in their decays, for instance through $\squark \rightarrow q
\ninoone$, while gluinos typically generate at least two jets, for instance through $\gluino\rightarrow q \bar{q} \ninoone$. Processes contributing to $\squark\squark$, $\squark\gluino$ and $\gluino\gluino$ final states therefore lead to events containing at least two, three or four jets, respectively. Cascade decays of heavy particles, as well as initial and final state radiation, tend to further increase the final state multiplicity. 

Inclusive analysis channels, labelled A to E and characterized by increasing minimum jet multiplicity from two to six, are therefore defined. In addition, the two-jet sample is divided into two channels, A and A$'$, using the ratio of  \met{}  to \meff{}, giving a total of six channels.  Channel A$'$ is designed to improve the sensitivity to models with small supersymmetric particle (`sparticle') mass splittings, where the presence of initial state radiation jets may allow signal events to be selected irrespective of the visibility of the sparticle decay products. The lower jet multiplicity channels focus on models characterized by squark pair production with short decay chains, while those requiring high jet multiplicity are optimized for gluino pair production and/or long cascade decay chains. The final limits are set using the channel with the best expected sensitivity for each hypothesis. The channels and signal regions (SRs) are summarized in Table~\ref{tab:srdefs}. The final selection criteria are defined without reference to collision data satisfying the criteria applied earlier in the selection.

The effective mass is defined to be the scalar sum of the transverse momenta of the leading $N$ jets in the event together with \met{}:

\begin{equation}\label{eq:meffdef}
\meff \equiv \sum_{i=1}^N \ourpt^{(i)} + \met .
\end{equation}

This general quantity is used to select events in two different ways, for which the specific values of $N$ used in the sum differ. Criteria are placed on the ratio of  \met{}  to $\meff$, in which context $N$ is defined to be the minimum number of jets used in the channel under consideration (for example $N=2$ for channel A). In Table~\ref{tab:srdefs}, where the number of jets used is explicitly notated, the expression $\meff$~($N$j) indicates the exact, exclusive, number of jets used. However, the final signal selection in all channels uses criteria on a more inclusive definition, $\meff({\rm incl.})$, for which the sum extends over all jets with $\ourpt>40$~\GeV. Requirements on $\meff$ and \met{}, which suppress the QCD multi-jet background, formed the basis of the previous ATLAS jets + \met{} + 0-lepton SUSY search~\cite{Aad:2011ib}. The same strategy is adopted in this analysis.

In Table~\ref{tab:srdefs}, $\ourdeltaphifull$ is the smallest of the
azimuthal separations between the missing momentum vector in the transverse plane, $\ourvecptmiss$, and the reconstructed jets. For channels A, A$'$ and B, the selection requires $\ourdeltaphifull>0.4$ radians using up to three leading jets. For the other channels an additional requirement $\ourdeltaphifull>0.2$ radians is applied to the remaining jets with $\ourpt>40$~\GeV. 
Requirements on $\ourdeltaphifull$ and $\MET/\meff$ are designed to reduce the background from multi-jet processes.

%%%%
\begin{table*}
  \begin{center}\renewcommand\arraystretch{1.4}
    \begin{tabular}{|l|c|c|c|c|c|c|}
      \hline
      \multirow{2}{*}{Requirement}      &\multicolumn{6}{|c|}{Channel} \\
\cline{2-7}
  & A &A$'$ & B & C & D & E \\
 \hline
 Trigger  &\multicolumn{6}{|c|} {Leading jet $\pt>75$~GeV (EM Scale) and $\met>45-55$~GeV} \\
 Lepton veto   &\multicolumn{6}{|c|} {No electron (muon) with $\pt>20$ (10)~GeV} \\
 \hline
\met [GeV] $>$&\multicolumn{6}{|c|}{ 160 }\\ \hline
$\pt(j_1)$ [GeV] $>$&\multicolumn{6}{|c|}{ 130 }\\ \hline
$\pt(j_2)$ [GeV] $>$&\multicolumn{6}{|c|}{ 60 }\\ \hline
$\pt(j_3)$ [GeV] $>$&-- &-- &60  &60  &60  &60 \\ \hline
$\pt(j_4)$ [GeV] $>$&-- &-- &-- &60  &60  &60 \\ \hline
$\pt(j_5)$ [GeV] $>$&-- &-- &-- &--  &40  &40 \\ \hline
$\pt(j_6)$ [GeV] $>$&-- &-- &-- &--  &--  &40 \\ \hline
$\ourdeltaphifull$ [rad] $>$ &\multicolumn{3}{|c|}{0.4 ($i=\{1,2,(3)\}$)} &\multicolumn{3}{|c|}{0.4 ($i=\{1,2,3\}$), 0.2 ($\pt>40$ GeV jets)}\\ \hline
$\met/\meff$ ($N$j) $>$ & 0.3 (2j) &0.4 (2j) &  0.25 (3j) &  0.25 (4j) &  0.2 (5j) & 0.15 (6j) \\ \hline
$ \meff({\rm incl.})$ [GeV] $>$ &  1900/1400/--  &--/1200/-- &  1900/--/-- &  1500/1200/900 &  1500/--/-- &  1400/1200/900 \\ 
\hline
\hline
\end{tabular}
\caption{\label{tab:srdefs} Criteria used to define each of the inclusive channels and streams in the analysis. The jets are ordered with the highest $\pt$ first. The variables used are defined in the text. The $\met/\meff$ selection in any $N$ jet channel uses a value of $\meff$ constructed from only the leading $N$ jets (indicated in parentheses).  However, the final $\meff$(incl.) selection, which is used to define the signal regions, includes all jets with $\pt>40$ GeV. The three $\meff$(incl.) values listed in the final row denote the `tight', `medium' and `loose' selections,  respectively, as used for the final SRs.}
  \end{center}
\end{table*}

SM background processes contribute to the event counts in
the signal regions. The dominant sources are: $W+$jets, $Z+$jets, top quark
pair, single top quark, diboson and multi-jet    
production.
The majority of the $W$+jets background is composed 
of $W\rightarrow \tau\nu$ events, or $W\rightarrow e\nu, \mu\nu$ events in which  
no electron or muon candidate is reconstructed.
The largest part of the $Z$+jets background comes from the irreducible
component in which $Z\rightarrow\nu\bar\nu$ decays generate large $\MET$. 
Top quark pair production followed by semi-leptonic decays, in particular $t \bar{t} \rightarrow b \bar{b} qq' \tau \nu$ with the $\tau$ lepton decaying hadronically, as well as single top quark events,
can also generate large $\MET$ and pass the jet
and lepton requirements at a non-negligible rate.
The multi-jet background in the signal regions is caused by
poor reconstruction of jet energies in the calorimeters leading to apparent
missing transverse momentum, as well as by neutrino production in semi-leptonic decays of heavy quarks. Extensive validation of the MC simulation against data has been performed for each of
these background sources and for a wide variety of control
regions (CRs).

Each of the six channels is used to construct between one and three signal regions with `tight', `medium' and/or `loose' $\meff({\rm incl.})$ selections, giving a total of 11 SRs.
In order to estimate the backgrounds in a consistent and robust fashion, five control regions are defined for each of the SRs, giving 55 CRs in total. 
Each ensemble of one SR and five CRs constitutes a different `stream' of the analysis. The CR selections are optimized to maintain adequate statistical weight, while minimizing as far as possible the systematic uncertainties arising from extrapolation to the SR, and any contamination from signal events. This is achieved by using kinematic selections that are as close as possible to the relevant SR, and making use of other event properties to create CR samples to measure the backgrounds. 

\begin{table*}
  \begin{center}\renewcommand\arraystretch{1.4}
    \begin{tabular}{| l | c | c | c |}
      \hline
      CR & SR Background &  CR process & CR selection \\ \hline
CR1a & $Z$+jets & $\gamma$+jets & Isolated photon \\
CR1b & $Z$+jets & $Z(\to\ell\ell)$+jets & $66 \GeV < m(\ell\ell) < 116 \GeV$\\
CR2 & Multi-jets & Multi-jets &  $\ourdeltaphifull <0.2$ rad\\
CR3 & $W(\to\ell\nu)$+jets & $W(\to\ell\nu)$+jets & 30 GeV $<m_{\rm T}(\ell,\met) < 100$ GeV, $b$-veto\\
CR4 & $t\bar{t}$ and single top & $t\bar{t}\to b\bar{b}qq'\ell\nu$ & 30 GeV $<m_{\rm T}(\ell,\met) < 100$ GeV, $b$-tag\phantom{o}\\
\hline
\end{tabular}
\caption{\label{tab:crdefs} Control regions used in the analysis: the main targeted background in the SR, the process used to model the background, and main CR selection(s) used to select this process are given.}
  \end{center}
\end{table*}

The CRs are listed in Table~\ref{tab:crdefs}. CR1a and CR1b are used to estimate the contribution of $Z(\rightarrow \nu\bar{\nu})$+jets background events to the SR by selecting samples of $\gamma$+jets and $Z(\rightarrow \ell\ell)$+jets events, respectively. The control region CR2 uses a reversed and tightened criterion on $\ourdeltaphifull$ for up to three selected leading jets (depending on channel) to produce a data sample enriched with multi-jet background events. Otherwise it uses identical kinematic selections to the SRs. CR3 and CR4 use respectively a $b$-jet veto or $b$-jet requirement together with a lepton+\met{} transverse mass ($m_\mathrm{T}$) requirement to select samples of  $W(\rightarrow \ell \nu)$+jets and semi-leptonic $t\bar{t}$ background events. Other selections are similar to those used to select the corresponding signal region, although in CR1b, CR3 and CR4 the requirements on $\ourdeltaphifull$ and $\met/\meff$ are omitted to maximize the number of events without introducing extrapolations in energy or jet multiplicity. 

The observed numbers of events in the CRs for each SR are used to generate internally consistent SM background estimates for the SR via a likelihood fit. This procedure enables CR correlations and contamination of the CRs by other SM processes and/or SUSY signal events to be taken into account. The same fit also allows the statistical significance of the observation in the SR with respect to the SM expectation to be determined.
The estimated number of background events for a given process, $N$\mbox{(SR, scaled)}, is given by
\begin{equation}
\label{eq:tf}
N\mathrm{(SR, scaled)}=N\mathrm{ (CR, obs)}
\times
 \left [\frac{N\mathrm{(SR, unscaled)}}{N\mathrm{(CR, unscaled)}}\right ] ,
 \end{equation}
where 
$N$(CR, obs) is the observed number of data events in the CR for the process, and $N$(SR, unscaled) and $N$(CR, unscaled) are estimates of the contributions from the process to the SR and CR, respectively, as described in Section~\ref{sec:back}. The ratio appearing in the square brackets in Eq.~(\ref{eq:tf}) is defined to be the transfer factor (TF). Similar equations containing inter-CR TFs enable the background estimates to be normalized coherently across all the CRs. The likelihood fit adjusts the predicted background components in the CRs and SRs using the TFs and the unscaled CR event counts as constraints, taking into account their uncertainties. The scaled values are output from the fit. 

The likelihood function for observing $n$ events in one of the channels (A--E, loose to tight) is the product of Poisson 
distributions, one for the signal region and one for each of the main control regions constraining 
the $Z$+jets (CR1a/b), multi-jets (CR2), $W$+jets (CR3) and $\ttbar$ (CR4) contributions, labelled $P_{\rm SR}$, $P_{\rm ZRa,b}$, 
$P_{\rm JR}$, $P_{\rm WR}$ and $P_{\rm TR}$ respectively, and of the PDFs constraining 
the systematic uncertainties $C_{\rm Syst}$:
\begin{eqnarray}
\label{stat_likelihood_0l}
  L(n|\mu,b,\theta)&=& P_{\rm SR} \cdot P_{\rm ZRa}  \cdot  P_{\rm ZRb}  \cdot P_{\rm JR} \cdot P_{\rm WR}  \cdot P_{\rm TR} \cr
  & & \phantom{}  \cdot C_{\rm Syst}(\theta) .
\end{eqnarray}

The total expected background is $b$. The expected means for the Poisson distributions are computed from the observed numbers of events in the control regions, using the TFs. The signal strength $\mu$ parameterizes the expected signal, with $\mu=1$ giving the full signal expected in a given model. The nuisance parameters ($\theta$) parameterize the systematic uncertainties, such as that on the integrated luminosity. 

The expected number of events in the signal region is denoted by $\lambda_S$, while
$\lambda_i$ denotes the expected number of events
in control region $i$.  These are expressed in terms of the fit
parameters $\mu$ and $b$ and an extrapolation matrix $C$ (connecting
background and signal regions) as follows:

\begin{eqnarray}
\lambda_{S}(\mu,b,\theta) &=& \mu \cdot C_{\rm{SR}\rightarrow {\rm SR}}(\theta)\cdot s \cr
                                               &+& \sum_{j} C_{j\rm{R} \rightarrow {\rm SR}}({\theta})\cdot b_{j}\,,\\
\lambda_{i}(\mu,b,\theta) &=& \mu \cdot C_{\rm{SR}\rightarrow i\rm{R}}({\theta})\cdot s \cr
                                              &+& \sum_{j} C_{j\rm{R}  \rightarrow i\rm{R}}({\theta})\cdot b_{j}\,,
\end{eqnarray}
where the index $j$ runs over the background control regions.  The observed number of signal events in the SR (CR$_{j\mathrm{R}}$) are s ($b_j$), respectively. The diagonal elements of the matrix are all unity by construction. The off-diagonal elements are the various TFs.
%The terms $C_{\rm{JR}\rightarrow \rm{JR}}$, $C_{\rm{WR}\rightarrow \rm{WR}}$,$C_{\rm{ZRa,b}\rightarrow \rm{ZRa,b}}$,
%$C_{\rm{TR}\rightarrow \rm{TR}}$ are by construction all equal to $1$.

This background estimation procedure requires the determination of the central expected values of the TFs for each SM process, together with their associated correlated and uncorrelated uncertainties, as described in Section~\ref{sec:back}. The multi-jet TFs are estimated using a data-driven technique, which applies a resolution function to well-measured multi-jet events in order to estimate the effect of mismeasurement on \met{} and other variables. 
The other TF estimates use fully simulated MC samples validated with data (see Section \ref{subsec:MC}). Some systematic uncertainties, for instance those arising from the jet energy scale (JES), or theoretical uncertainties in MC simulation cross sections, largely cancel when calculating the event count ratios constituting the TFs. 

The result of the likelihood fit for each stream includes a set of background estimates and uncertainties for the SR together with a p-value giving the probability for the hypothesis that the observed SR event count is compatible with background alone. 
Conservative assumptions are made about the migration of SUSY signal events between regions. When seeking an excess due to a signal in a particular SR,  it is assumed that the signal contributes only to the SR, i.e. the SUSY TFs are all set to zero, giving no contribution from signal in the CRs. 
If no excess is observed, then limits are set within specific SUSY parameter spaces, taking into account theoretical and experimental uncertainties on the SUSY production cross section and kinematic distributions.
Exclusion limits are obtained using a likelihood test. This compares the observed event rates in the signal regions with the fitted background expectation and expected signal contributions, for various signal hypotheses. Since the signal hypothesis for any specific model predicts the SUSY TFs, these exclusion limits do allow for signal contamination in the CRs.

\section{Data and Simulated Samples}
\label{sec:samples}
\subsection{Proton-Proton Collision Data Sample}

The data used in this analysis were taken in 2011 with the LHC operating at a center-of-mass energy of 7~\TeV. Over this period the peak instantaneous luminosity increased from $1.3\times10^{30}$  to $3.7\times10^{33}$~cm$^{-2}$s$^{-1}$ and the peak mean number of interactions per bunch crossing increased from 2 to 12. 
Application of beam, detector and data-quality requirements resulted in
a total integrated luminosity of \ourintlumi~~\cite{Aad:2011dr,ATLAS-CONF-2011-116}. The precision of the luminosity measurement is 3.9\%.
The trigger used is described in Section \ref{sec:trigger}. 

\subsection{Monte Carlo Samples}
\label{subsec:MC}

MC samples are used to develop the analysis, optimize the selections, determine the transfer factors used to estimate the $W$+jets, $Z$+jets and top quark production backgrounds, and to assess sensitivity to specific SUSY signal models.  
Samples of simulated multi-jet events are generated with \pythia~\cite{Sjostrand:2006za}, using the {\tt MRST2007LO*} modified leading-order parton distribution functions (PDFs)~\cite{Sherstnev:2007nd}, for use in the data-driven background estimates.
Production of top quark pairs, including accompanying jets, is simulated with \alpgen~\cite{Mangano:2002ea} and the {\tt CTEQ6L1} \cite{Pumplin:2002vw} PDF set,
with a top quark mass of 172.5~GeV. Samples of $W$ and $Z/\gamma^*$ events with accompanying jets are also produced
with \alpgen.
Diboson ($WW$, $WZ$, $ZZ$, $W\gamma^*$) production is simulated with \sherpa~\cite{Gleisberg:2008ta}.
Single top quark production is simulated with {\tt AcerMC} \cite{Kersevan:2004yg}.
Fragmentation and hadronization for the \alpgen~samples is
performed with \herwig~\cite{Corcella:2000bw,herwig65long}, using
\jimmy~\cite{Butterworth:1996zw} for the underlying event.
For the $\gamma$+jet estimates of the $Z(\to\nu\bar{\nu})$+jets backgrounds, photon and $Z$ events are also both produced using \sherpa{} for consistency checks of the \alpgen{} results.

SUSY signal samples are generated with {\tt Herwig++}~\cite{Bahr:2008pv} or {\tt MadGraph}/{\pythia}\cite{Alwall:2007st,madgraph1,Sjostrand:2006za}. 
Signal cross sections are calculated to next-to-leading order in the strong coupling constant, including the resummation of soft gluon emission at next-to-leading-logarithmic accuracy (NLO+NLL)~\cite{Beenakker:1996ch,Kulesza:2008jb,Kulesza:2009kq,Beenakker:2009ha,Beenakker:2011fu}~\cite{susyxsec}. The nominal cross section and the uncertainty are taken from an ensemble of cross section predictions using different PDF sets and factorisation and renormalisation scales, as described in Ref.~\cite{Kramer:2012bx}.

The MC samples are generated using the same parameter set as Refs.~\cite{ATL-PHYS-PUB-2011-009,ATL-PHYS-PUB-2011-014,ATLAS:1303025} and passed through the ATLAS detector simulation \cite{:2010wqa} based on {\tt GEANT4} \cite{Agostinelli:2002hh}. Differing pile-up (multiple proton-proton interactions in a given event)
conditions as a function of the LHC instantaneous luminosity are taken into account by overlaying simulated
minimum-bias events onto the hard-scattering process and reweighting them according to the expected mean
number of interactions per LHC bunch crossing.

\section{Trigger Selections}
\label{sec:trigger}
The baseline triggers for the signal region event selection in the 2011 analysis use jets and  \met \cite{Aad:2012xs, ATLAS-CONF-2011-072}. The jet and $\met$ trigger
required events to contain a leading jet with a transverse momentum ($\ourpt$), measured at the electromagnetic energy scale~\cite{emscale}, above 75~\GeV{} and significant missing transverse momentum. The detailed
trigger specification, including the value of the $\met$ threshold, varied throughout the data-taking period, partly as a consequence of the rapidly
increasing LHC luminosity. The trigger threshold on the missing transverse momentum increased from 45 GeV at the start of the data-taking period, to 55 GeV at the end.
The trigger reached its full efficiency of $>$ 98\% for events with  a reconstructed jet with $\ourpt$ exceeding 130~\GeV{} and more than 160~\GeV{} of missing transverse momentum. 
Trigger efficiencies are extracted using a sample selected by a looser trigger,
taking into account correlations, i.e. correcting for the efficiency
of the looser trigger. Prescaled single-jet triggers, which acquired fixed fractions of the data, are used for the trigger efficiency study. 

A second study verifies that the efficiency of the baseline trigger becomes maximal at the values quoted above. 
The efficiencies are determined with an independent sample of events expected to possess \met{} generated by neutrinos. 
A sample triggered by electron candidates is used, where jets from electrons reconstructed with tight selection criteria are discarded. This trigger selected mostly
$W\to e\nu$ events with jets and ran unprescaled, thus
providing a large number of events.

\section{Object Reconstruction}
\label{sec:objects}

The event reconstruction algorithms create the physics objects used in this analysis: electrons, muons, jets, photons  and $b$-jets. Once these objects are defined, the overall missing transverse momentum can be calculated. A failure in the calorimeter electronics created a small dead region ($0<\eta<1.4$, $-0.8<\phi<-0.6$) in the second and third layers of the electromagnetic calorimeter, which affected energy measurements in about 20\% of the data sample. Any event with a jet that is inside the affected region and that is expected on the basis of shower shape to potentially contribute significantly to the \met{} is removed from the sample to avoid fake signals~\cite{smartveto}. The energies of jets inside the affected region which are not expected to create \met{} are corrected using the functioning calorimeter layers. 

Jet candidates are reconstructed using the
anti-$k_t$ jet clustering algorithm~\cite{Cacciari:2008gp,Cacciari:2005hq} with a
radius parameter of $0.4$. The inputs to this algorithm are clusters~\cite{Lampl:2008}
of calorimeter cells seeded by those with energy significantly above
the measured noise. Jet momenta are constructed by performing a
four-vector sum over these cell clusters, measured at the electromagnetic scale, treating each as an
$(E,\vec{p})$ four-vector with zero mass.  The jet energies are corrected for
the effects of calorimeter non-compensation and inhomogeneities by
using $\ourpt$- and $\eta$-dependent calibration factors
derived from MC simulation and validated with extensive
test-beam and collision-data studies \cite{atlas-jes-paper2011}. 
Only jet candidates with $\ourpt > 20$~\GeV{} 
are subsequently retained.

Electron candidates are required to have $\ourpt > 20$~\GeV{} and $|\eta| <
2.47$, and to pass the `medium' electron shower shape and track selection criteria described in
Ref.~\cite{Aad:2011mk}.
Muon candidates \cite{ATLAS-CONF-2011-046,ATLAS-CONF-2011-063} are required to have matching tracks in the inner detector and muon spectrometer with $\ourpt > 10$~GeV and $|\eta| <
2.4$.  

Following the steps above, overlaps between candidate jets with $|\eta|<2.8$ and leptons are resolved as follows:
first, any such jet candidate lying within a distance $\Delta
R\equiv\sqrt{(\Delta\eta)^2+(\Delta\phi)^2}=0.2$ ($\phi$ measured in radians) of an electron is discarded;
then any lepton candidate remaining within a distance
$\Delta R =0.4$ of any surviving jet candidate is discarded. 
The first requirement prevents energy deposits from being interpreted as both jets and electrons. The second ensures that leptons produced within jets are not used to veto the event during the selection described in Section~\ref{sec:evsel}.

The measurement of the missing transverse momentum two-vector
$\ourvecptmiss$ is based on
the transverse momenta of all remaining jet and lepton candidates and all
calorimeter clusters not associated with such objects.  Following this step, all jet candidates with $|\eta|>2.8$ are discarded, owing to their lower precision. 
Thereafter, the remaining lepton and jet candidates are considered
``reconstructed'', and the term ``candidate'' is dropped.

Photons are identified with the same selection criteria as used in the ATLAS prompt photon cross section analysis \cite{ATLAS:2012ar}, where an isolated photon passing the 
tight photon identification criteria is required. Jets are classified as $b$-jets using a neural network algorithm, which takes as inputs the impact parameter measurements and the topological structure of $b$-quark decays, as described in Refs.~\cite{ATLAS-CONF-2011-102,ATLAS-CONF-2012-043}.

\section{Removal of Non-Collision Backgrounds}
\label{sec:evclean}

Non-collision backgrounds are produced predominantly by noise sources in the calorimeters, cosmic ray events and beam collisions with residual gas in the beam-pipe (beam-gas events). The requirement of a vertex near the nominal interaction point with at least five associated tracks is effective at suppressing these backgrounds. Further criteria are applied which require that the fractional energy deposited in each calorimeter layer, and in any cells with known quality problems, is consistent with that expected from beam-beam events. In addition, the energy observed in charged particle tracks associated with the calorimeter cluster, and the timing of the energy depositions in calorimeter cells with respect to the beam-crossing time are checked~\cite{atlas-jes-paper2011}. Following these selections, the remaining background is estimated by using the observed time distribution of the leading jets with respect to the bunch crossing, to create a background dominated control region. The non-collision background is found to be negligible in all of the SRs and CRs used.

\section{Event Selection}
\label{sec:evsel}

Following the object reconstruction and event cleaning described above, a lepton veto is applied to reject $W(\rightarrow\ell\nu)$+jets and leptonic $t\bar{t}$ events in which neutrinos generate the $\met$ signature. The lepton $\pt$ threshold used in the veto is set at 20 (10) GeV for electrons (muons) to ensure that selected events correspond to a phase space region in which the
veto efficiency is well understood. 

The signal regions are then defined by the kinematic selections given in Table~\ref{tab:srdefs}. 
Requirements on the transverse momenta of additional jets select  inclusive 2-, 3-, 4-, 5- and 6-jet events in channels A/A$'$, B, C, D and E  respectively. The jet $\pt$ thresholds for the leading up to four jets are set at 60~\GeV{} in order to minimize the impact of pile-up on selection efficiency and improve background rejection. 

Removing events with a small angle in the transverse plane ($\Delta\phi$) between jets and $\met$ suppresses multi-jet background in which mismeasurement of jet energy
generates fake missing transverse momentum along the jet direction. 
For channels A, A$'$ and B a requirement $\Delta\phi> 0.4$ radians is applied to the leading (up to) three selected jets with $\pt$ $>$ 40~\GeV{}, before the final SR selection, to minimize loss of signal efficiency. For the other channels this requirement is augmented by a looser requirement that $\Delta\phi>0.2$ radians for all remaining selected jets with $\pt$ $>$ 40~\GeV. 

Multi-jet background is further suppressed by requiring that the $\met$ exceeds a specific fraction of the effective mass of the event, $\meff$. Coupled with the explicit requirement on $\meff$(incl.) discussed below this equates to a hard selection on $\met$. The $\met$/$\meff$ value used decreases with increasing jet multiplicity because the typical $\met$ of SUSY signal events is inversely correlated with jet multiplicity due to phase-space limitations. This is because additional jets in a SUSY decay chain increase the probability that the lightest SUSY particle (LSP) will be produced with low momentum through effective multi-body decays. Small mass splittings can also lead to low $\met$. The multi-jet cross section is also suppressed at higher jet multiplicities, allowing the $\met$ requirement to be loosened.

Finally, the signal regions are defined by criteria on $\meff$(incl.) which select events with hard kinematics in order to provide strong suppression of all SM background processes. Up to three $\meff$(incl.) values are specified per channel, corresponding to distinct signal regions `tight', `medium' and `loose', in which the final event samples are counted.  

\begin{figure}
\vspace*{-0.5cm}
\begin{center}
\includegraphics[height=0.43\textwidth]{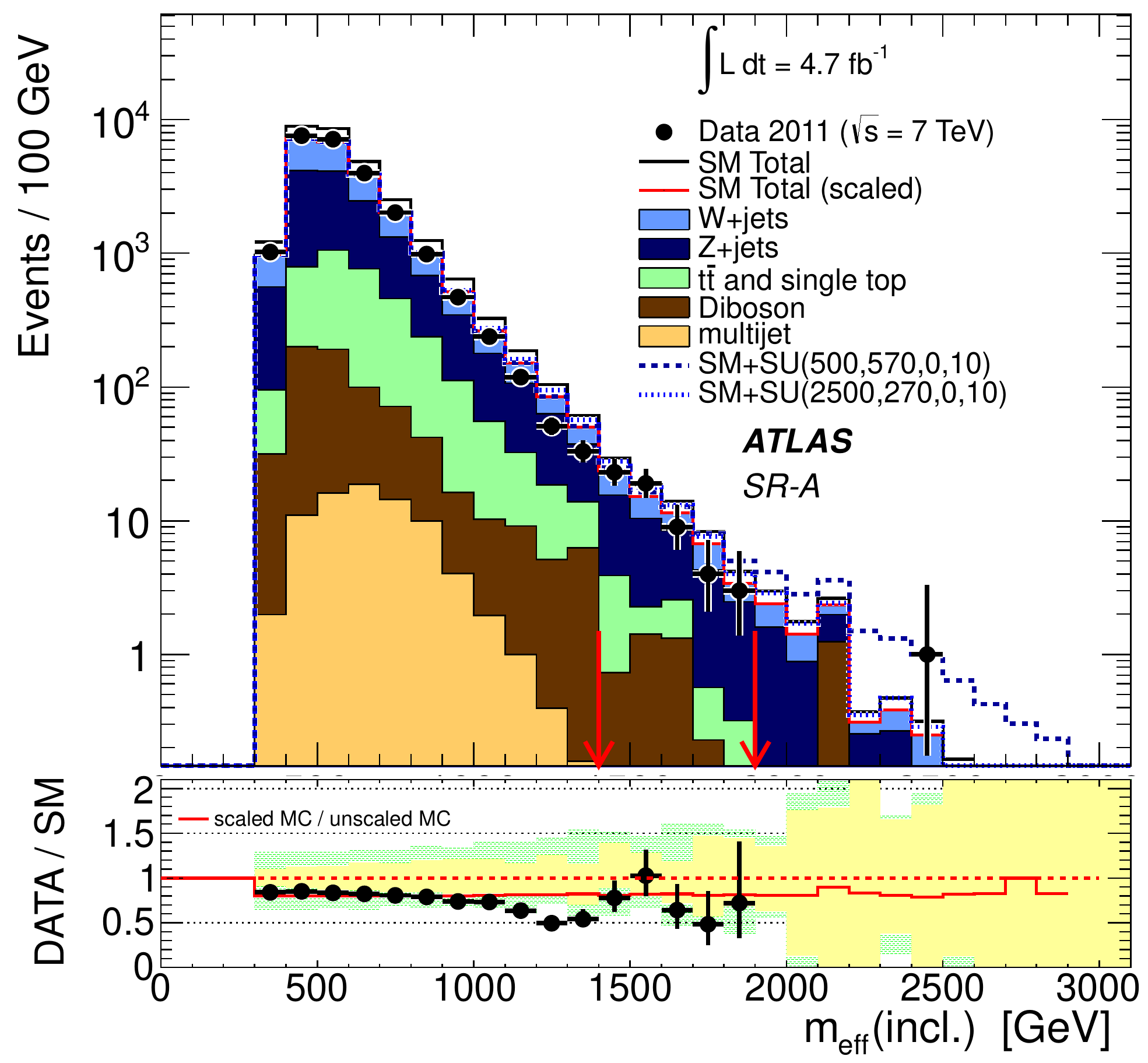}
\end{center}
\vspace*{-0.5cm}
\caption{Observed $\meff({\rm incl.})$ distribution for channel A.  
In the top panel, the histograms show the SM background expectations, both before (black open histogram) and after (medium (red) open histogram) use of a fit to scale the expectations to CR observations. This fit is applied to illustrate the SR+CR fitting technique used in the main analysis.
Before scaling, the MC simulation expectations are normalized to luminosity. 
The multi-jet background is estimated using the jet smearing method described in the text.
After scaling, the $W$+jets, $Z$+jets and $t\bar{t}$
and single top quark
and multi-jet distributions (denoted by full histograms) are 
normalized to data in corresponding control regions over the full $\meff$ range.  
Two MSUGRA/CMSSM benchmark model points with $m_{\rm 0}$=500~\GeV, $m_{\rm 1/2}$=570\ GeV, $A_{\rm 0}$=0, $\tan\beta$=10 and $\mu>$ 0 and with  $m_{\rm 0}$=2500~\GeV, $m_{\rm 1/2}$=270\ GeV, $A_{\rm 0}$=0, $\tan\beta$=10 and $\mu>$ 0, illustrating different topologies, are also shown. These points lie just beyond the reach of the previous analysis \cite{Aad:2011ib}. The arrows indicate the locations of the lower edges of the two signal regions.
The bottom panel shows the fractional deviation of the data from the total unscaled background estimate (black points), together with the fractional deviation of the total scaled background estimate from the total unscaled background estimate (medium (red) line). 
The light (yellow) band shows the combined experimental uncertainties on the unscaled background estimates from jet energy scale, jet energy resolution, the effect of pile-up, the treatment of energy outside of reconstructed jets and MC simulation sample size. 
The medium (green) band includes also the total theoretical uncertainties.}
\label{fig:sra}
\end{figure}

\begin{figure}
\begin{center}
\includegraphics[height=0.43\textwidth]{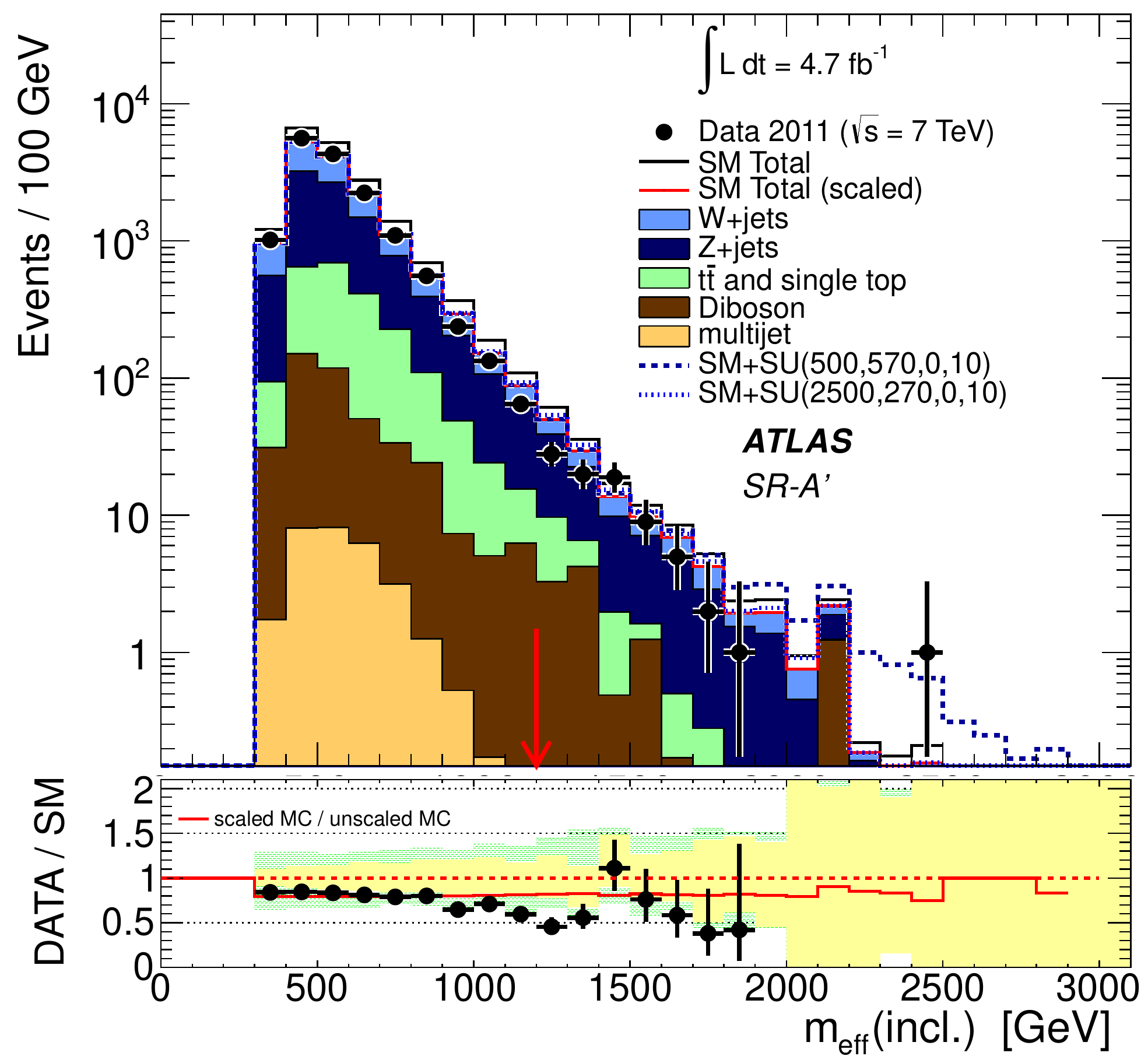}
\end{center}
\vspace{-0.5cm}
\caption{\label{fig:srap}Observed $\meff({\rm incl.})$ distribution for channel A$'$, as for Fig.~\ref{fig:sra}.}
\end{figure}

\begin{figure}
\vspace*{-0.5cm}
\begin{center}
\includegraphics[height=0.43\textwidth]{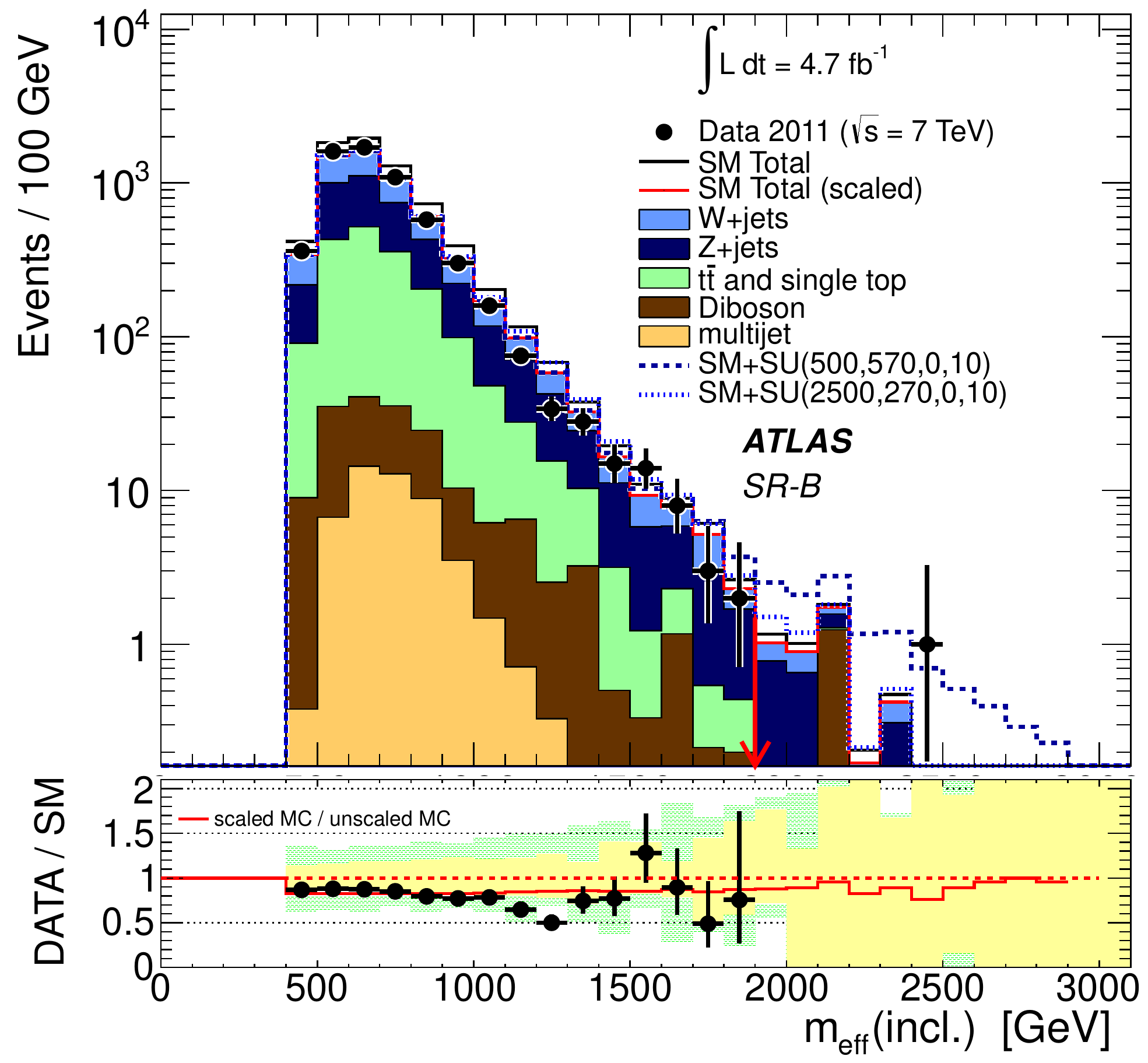}
\end{center}
\vspace*{-0.5cm}
\caption{\label{fig:srb}Observed $\meff({\rm incl.})$ distribution for channel B, as for Fig.~\ref{fig:sra}.}
\end{figure}

\begin{figure}
\begin{center}
\includegraphics[height=0.43\textwidth]{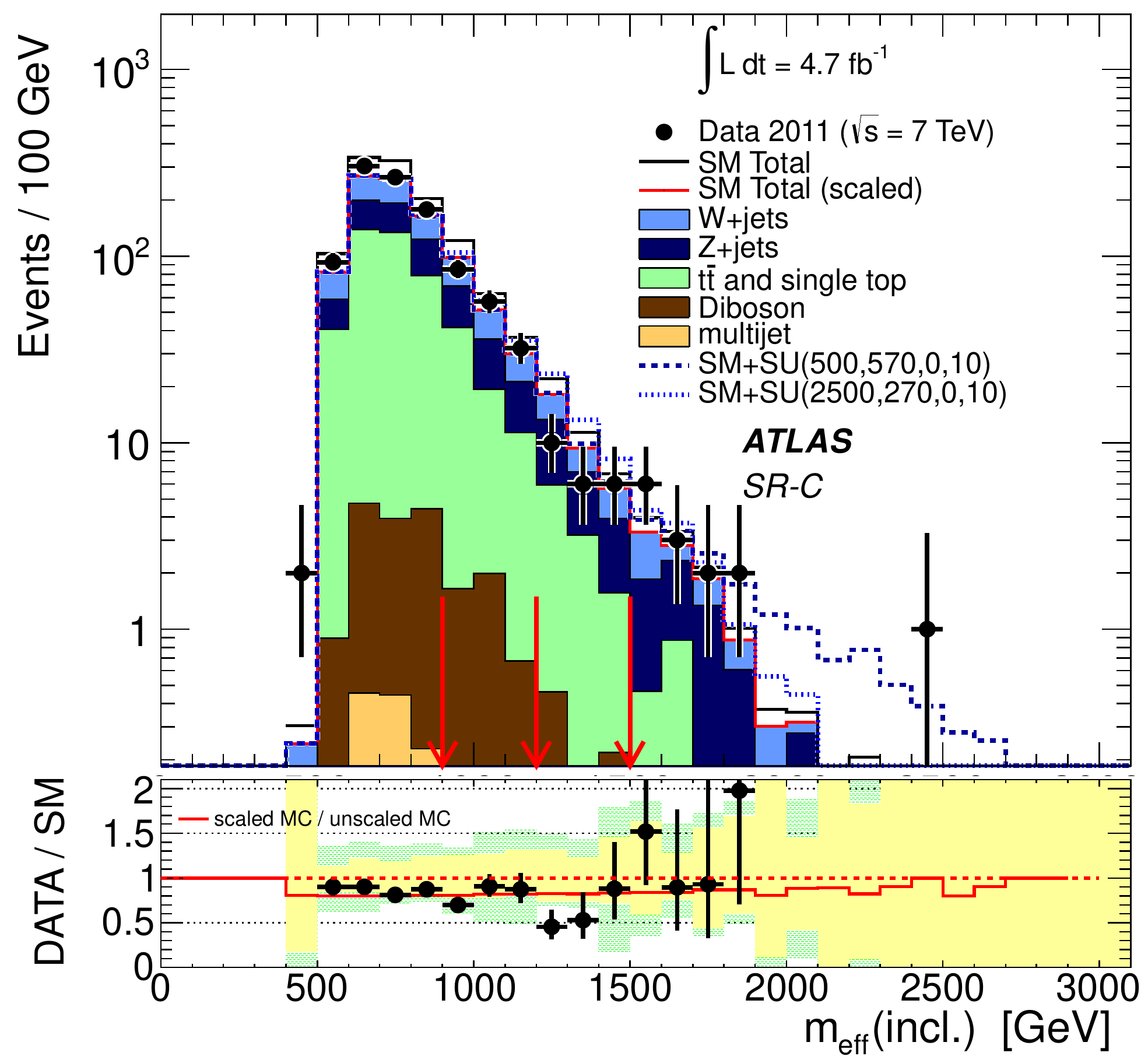}
\end{center}
\vspace*{-0.5cm}
\caption{\label{fig:src}Observed $\meff({\rm incl.})$ distribution for channel C, as for Fig.~\ref{fig:sra}.}
\end{figure}

\begin{figure}
\vspace*{-0.5cm}
\begin{center}
\includegraphics[height=0.43\textwidth]{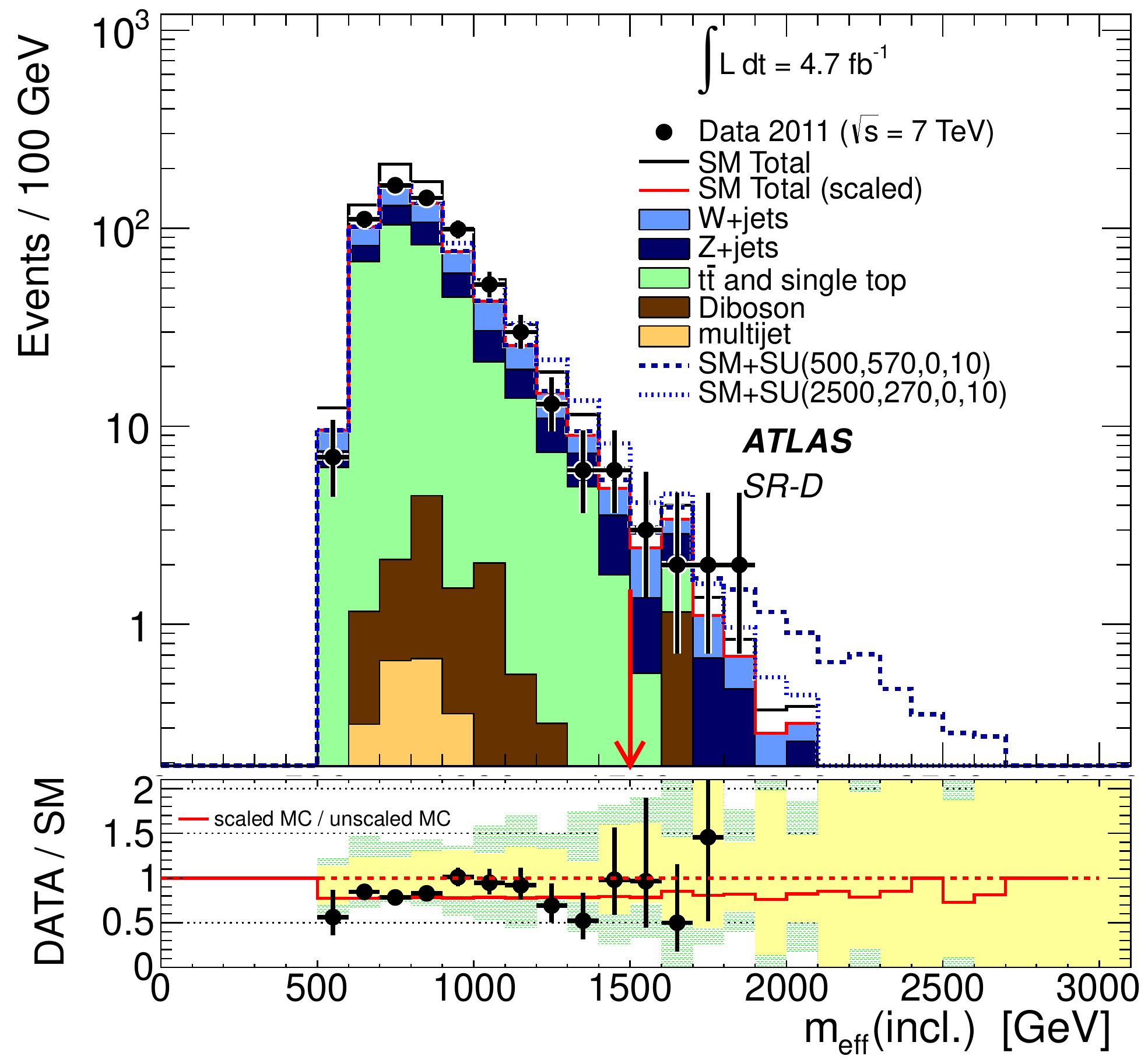}
\end{center}
\vspace*{-0.5cm}
\caption{\label{fig:srd}Observed $\meff({\rm incl.})$ distribution for channel D, as for Fig.~\ref{fig:sra}.}
\end{figure}

\begin{figure}
\begin{center}
\includegraphics[height=0.43\textwidth]{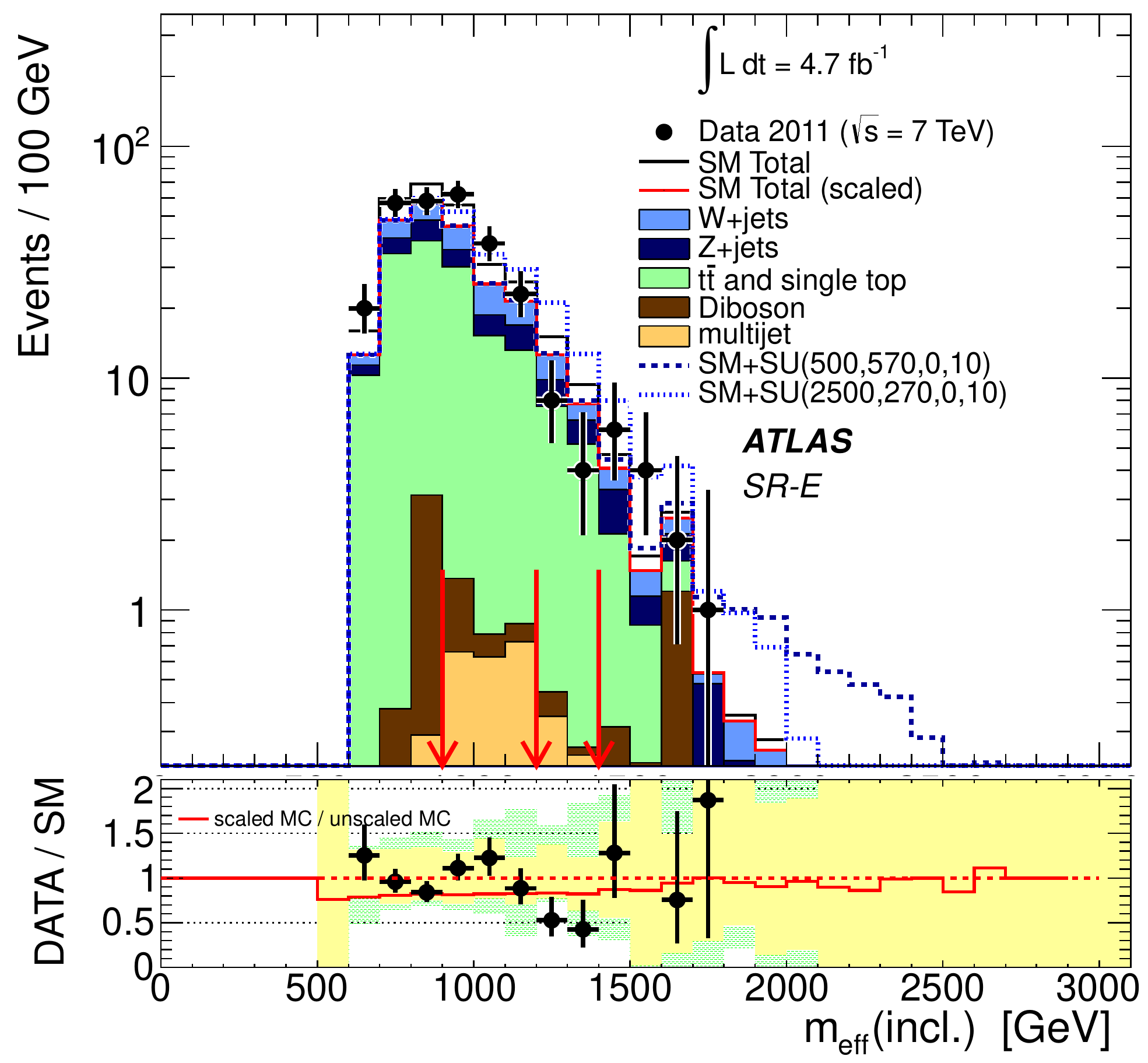}
\end{center}
\vspace*{-0.5cm}
\caption{\label{fig:sre}Observed $\meff({\rm incl.})$ distribution for channel E, as for Fig.~\ref{fig:sra}.}
\end{figure}

Table~\ref{tab:SRev} lists the number of data events passing each of the SR selections. The distributions of $\meff$(incl.) (prior to the final $\meff$(incl.) selections) for each channel for data and SM backgrounds are shown in Figs.~\ref{fig:sra}--\ref{fig:sre}. Details of the CR selections, and the methods used to obtain the background estimates follow in Section~\ref{sec:back}. The information is used in Section~\ref{sec:stats} to produce the final results.

\begin{sidewaystable*}
\begin{center}\renewcommand\arraystretch{1.4}
\begin{tabular}{|c|c|c|c|c|c|c|}
\hline
\multirow{2}{*}{Process}     &\multicolumn{6}{|c|}{Signal Region} \\ 
\cline{2-7}
                                         & SR-C loose                  & SR-E loose              & SR-A medium                & SR-A$'$ medium           & SR-C medium                      & SR-E medium \\
\hline
\hline
$t\bar{t}$+ single top  &  $74\pm 14$ (75) &  $73\pm 25$ (68) &  $6.8\pm 4.7$ (5.3) &  $11\pm 4$ (10) &  $13\pm 5$ (11) &  $19\pm 6$ (15) \\
$Z$+jets  &  $71\pm 19$ (78) &  $21\pm 7$ (17) &  $32\pm 9$ (44) &  $66\pm 18$ (88) &  $16\pm 5$ (20) &  $8.4\pm 3.2$ (5.6) \\
 $W$+jets &  $61\pm 11$ (61) &  $23\pm 13$ (23) &  $19\pm 5$ (21) &  $25\pm 5$ (30) &  $7.7\pm 3.0$ (11) &  $6.2\pm 2.6$ (4.7) \\
Multi-jets   &  $0.9\pm 1.2$ (0.8) &  $8.4\pm 7.3$ (25) &  $0.1\pm 0.3$ (0.2) &  $0.0\pm 0.1$ (0.5) &  $0.03\pm 0.05$ (0.03) &  $1.4\pm 1.2$ (2.7) \\
Di-bosons &  $7.9\pm 4.0$ (7.9) &  $4.2\pm 2.1$ (4.2) &  $7.3\pm 3.7$ (7.5) &  $14\pm 7$ (16) &  $1.7\pm 0.9$ (1.7) &  $2.7\pm 1.3$ (2.7) \\
Total  & $214\pm 8\pm 22$ & $129\pm 8\pm 30$ & $65\pm 4\pm 11$ & $116\pm 5\pm 19$ & $39\pm 3\pm 7$ & $38\pm 4\pm 5$ \\
Data  & 210 & 148 & 59 & 85 & 36 & 25 \\
Local p-value (Gauss. $\sigma$)   & 0.56($-$0.15) & 0.21(0.81) & 0.66($-$0.40) & 0.90($-$1.3) & 0.61($-$0.27) & 0.87($-$1.1) \\
Upper limit on $N_{\rm BSM}$  & 51($55 ^{\uparrow42 }_{\downarrow76 }$) & 77($67 ^{\uparrow49 }_{\downarrow91 }$) & 24($28 ^{\uparrow20 }_{\downarrow39 }$) & 28($42 ^{\uparrow31 }_{\downarrow58 }$) & 17($19 ^{\uparrow14 }_{\downarrow26 }$) & 11($16 ^{\uparrow12 }_{\downarrow23 }$) \\
Upper limit on $\sigma$ (fb)    & 11($12 ^{\uparrow8.8 }_{\downarrow16 }$) & 16($14 ^{\uparrow10 }_{\downarrow19 }$) & 5.1($5.9 ^{\uparrow4.3 }_{\downarrow8.3 }$) & 6.0($8.9 ^{\uparrow6.6 }_{\downarrow12 }$) & 3.6($4 ^{\uparrow2.9 }_{\downarrow5.6 }$) & 2.2($3.4 ^{\uparrow2.5 }_{\downarrow4.8 }$) \\
  \hline
 \end{tabular}
\begin{tabular}{|c|c|c|c|c|c|}
\hline
\multirow{2}{*}{Process}     &\multicolumn{5}{|c|}{Signal Region} \\ 
\cline{2-6}
& SR-A tight & SR-B tight & SR-C tight & SR-D tight & SR-E tight \\
\hline
\hline
$t\bar{t}$+ single top  &  $0.2\pm 0.2$ (0.1) &  $0.3\pm 0.3$ (0.2) &  $2.0\pm 1.5$ (1.2) &  $2.4\pm 1.7$ (1.4) &  $4.2\pm 4.7$ (3.0) \\
$Z$+jets  &  $3.3\pm 1.5$ (4.0) &  $2.0\pm 1.3$ (2.1) &  $2.0\pm 1.0$ (5.6) &  $0.9\pm 0.6$ (3.4) &  $3.4\pm 1.6$ (2.3) \\
 $W$+jets &  $2.2\pm 1.0$ (1.9) &  $1.0\pm 0.6$ (0.8) &  $1.5\pm 1.3$ (2.7) &  $2.4\pm 1.4$ (2.5) &  $2.8\pm 1.9$ (1.5) \\
Multi-jets   &  $0.00\pm 0.02$ (0.01) &  $0.00\pm 0.07$ (0.02) &  $0.00\pm 0.03$ (0.01) &  $0.0\pm 0.3$ (0.1) &  $0.5\pm 0.4$ (0.9) \\
Di-bosons &  $1.8\pm 0.9$ (2.0) &  $1.8\pm 0.9$ (1.9) &  $0.5\pm 0.3$ (0.5) &  $2.2\pm 1.1$ (2.2) &  $2.5\pm 1.3$ (2.5) \\
Total  & $7.4\pm 1.3\pm 1.9$ & $5.0\pm 0.9\pm 1.7$ & $6.0\pm 1.0\pm 2.0$ & $7.8\pm 1.0\pm 2.4$ & $13\pm 2\pm 6$ \\
Data  & 1 & 1 & 14 & 9 & 13 \\
Local p-value (Gauss. $\sigma$)   & 0.98($-$2.1) & 0.96($-$1.7) & 0.016(2.1) & 0.29(0.55) & 0.45(0.14) \\
Upper limit on $N_{\rm BSM}$  & 3.1($6.4 ^{\uparrow4.5 }_{\downarrow9.4 }$) & 3.0($5.6 ^{\uparrow3.9 }_{\downarrow8.3 }$) & 16($9.5 ^{\uparrow6.9 }_{\downarrow14 }$) & 9.6($8.5 ^{\uparrow6.1 }_{\downarrow12 }$) & 12($12 ^{\uparrow8.4 }_{\downarrow17 }$) \\
Upper limit on $\sigma$ (fb)    & 0.66($1.4 ^{\uparrow0.96 }_{\downarrow2.0 }$) & 0.64($1.2 ^{\uparrow0.83 }_{\downarrow1.8 }$) & 3.4($2.0 ^{\uparrow1.5 }_{\downarrow2.9 }$) & 2.0($1.8 ^{\uparrow1.3 }_{\downarrow2.6 }$) & 2.5($2.5 ^{\uparrow1.8 }_{\downarrow3.5 }$) \\
  \hline
 \end{tabular}
\caption{\label{tab:SRev} 
Observed numbers of events in data and fitted background components in each SR. For the total
background estimates, the quoted uncertainties give the statistical (MC simulation and CR combined) and systematic
uncertainties respectively. For the individual background components, the total
uncertainties are given, while the values in parenthesis indicate the pre-fit predictions. The predictions for $W$+jets, $Z$+jets and $\ttbar$ plus single top quark are from \alpgen{} and are normalized to luminosity. 
In the case of the multi-jet background, the pre-fit values are from the data-driven method, normalized at low \meff. The di-boson background is estimated with MC simulation normalized to luminosity.
The p-values give the probability of the observation being consistent with the estimated background, and the `Gauss. $\sigma$' values give the number of standard deviations in a Gaussian approximation, evaluated for a single observation at a time. The last two rows show the upper limits on the excess number of events, and the excess cross section, above that expected from the SM. The observed upper limit is followed in brackets by the expected limit, with the super- and sub-scripts showing the expectation from $\pm1\sigma$ changes in the background (denoted by $\uparrow$ and $\downarrow$ respectively).}
 \end{center}
 \end{sidewaystable*}

\section{Background Estimation}
\label{sec:back}

\subsection{Introduction}
The $Z(\to \nu\bar{\nu})$+jets process constitutes the dominant irreducible background in this analysis. It is estimated using control regions enriched in related processes with similar kinematics: events with isolated photons and jets \cite{Ask:2011xf} (CR1a, Section~\ref{gammajet})
and $Z(\to ee/\mu\mu)$+jets events (CR1b, Section~\ref{zjet}). The reconstructed momentum of the photon or the lepton-pair system is added to $\ourvecptmiss$ to obtain an estimate of the $\MET$ observed in $Z(\to \nu\bar{\nu})$+jets events. The predictions from both control regions are found to be in good agreement, and both are used in the final fit. The small additional background contributions from $Z(\to ee/\mu\mu/\tau\tau)$ decays in which the leptons are misidentified or unreconstructed, and from misidentified photon events, are estimated using the same control regions with appropriate transfer factors. 
The TF for CR1a estimates $Z(\to\nu\bar\nu)$+jets in the SR, and is corrected to give an estimate of $Z$+jets in the SR by multiplying by the ratio of $Z$+jets events to $Z(\to\nu\bar{\nu})$+jets events derived from MC simulation. In the case of CR1b the TF is calculated between $Z(\to ee/\mu\mu/\tau\tau)$+jets in the CR and $Z(\to \nu\bar{\nu}/ee/\mu\mu/\tau\tau)$+jets in the SR. Thus both methods ultimately provide an estimate of the total $Z$+jets background in the SR.

The backgrounds from multi-jet processes are estimated using a data-driven technique based upon the convolution of jets in a low $\MET$ data sample with jet response functions derived from multi-jet dominated data control regions (Section~\ref{subsec:qcd}).  
Those from $W$+jets and top quark processes are derived from MC simulation (Section~\ref{subsec:wtop}).

For each stream a likelihood fit is performed to the observed event counts in the five CRs, taking into account correlations in the systematic uncertainties in the transfer factors.

\subsection{ \texorpdfstring{$Z$}{Z}+jets estimate using a \texorpdfstring{$\gamma$}{gamma} + jets control region}
\label{gammajet}

The magnitude of the irreducible background from $Z (\rightarrow \nu\bar{\nu})$+jets 
events in the SRs can be estimated using $\gamma$ + jets data. When the vector boson $\pT$ is large, as 
required by the SR selections, the $Z$ and $\gamma$ cross sections differ mainly 
by their coupling constants with respect to quarks. For this reason the cross 
section ratio,
\begin{equation}
R_{Z/\gamma} = \frac{d \sigma(Z + {\rm jets}) / d\pT}{d \sigma(\gamma +{\rm  jets})/ d\pT}
\label{eq:gj:1}
\end{equation}
can be used to translate the observed number of photon events in the CR into an estimate of the number 
of $Z$ events in the SR, taking into account the leptonic branching ratios of the $Z$ and other effects. The ratio is expected to be robust with respect to both
theoretical uncertainties and experimental effects, related to, for example, 
jet reconstruction, which would be similar for both processes and therefore cancel in the ratio. 

The method uses photon events which are selected in two steps. The first aims to 
select a photon event sample where the efficiency and the background contamination 
are well known. 
The SR selections are then applied to these photon 
events, having added the photon $\pt$ to the $\met$ of the event to reproduce the $\met$ observed in $Z(\rightarrow \nu\bar{\nu})$ background events. The SR selections consist primarily of requirements on the jets and \met\ in the event, 
which directly or indirectly, due to the $p_{\rm T}$ recoil, impose kinematic constraints
on the vector boson, i.e. the $Z$ or photon. 

\begin{figure*}
\begin{center}
\subfigure[]{\label{fig:zfg:crA} \includegraphics[width=0.463\textwidth]{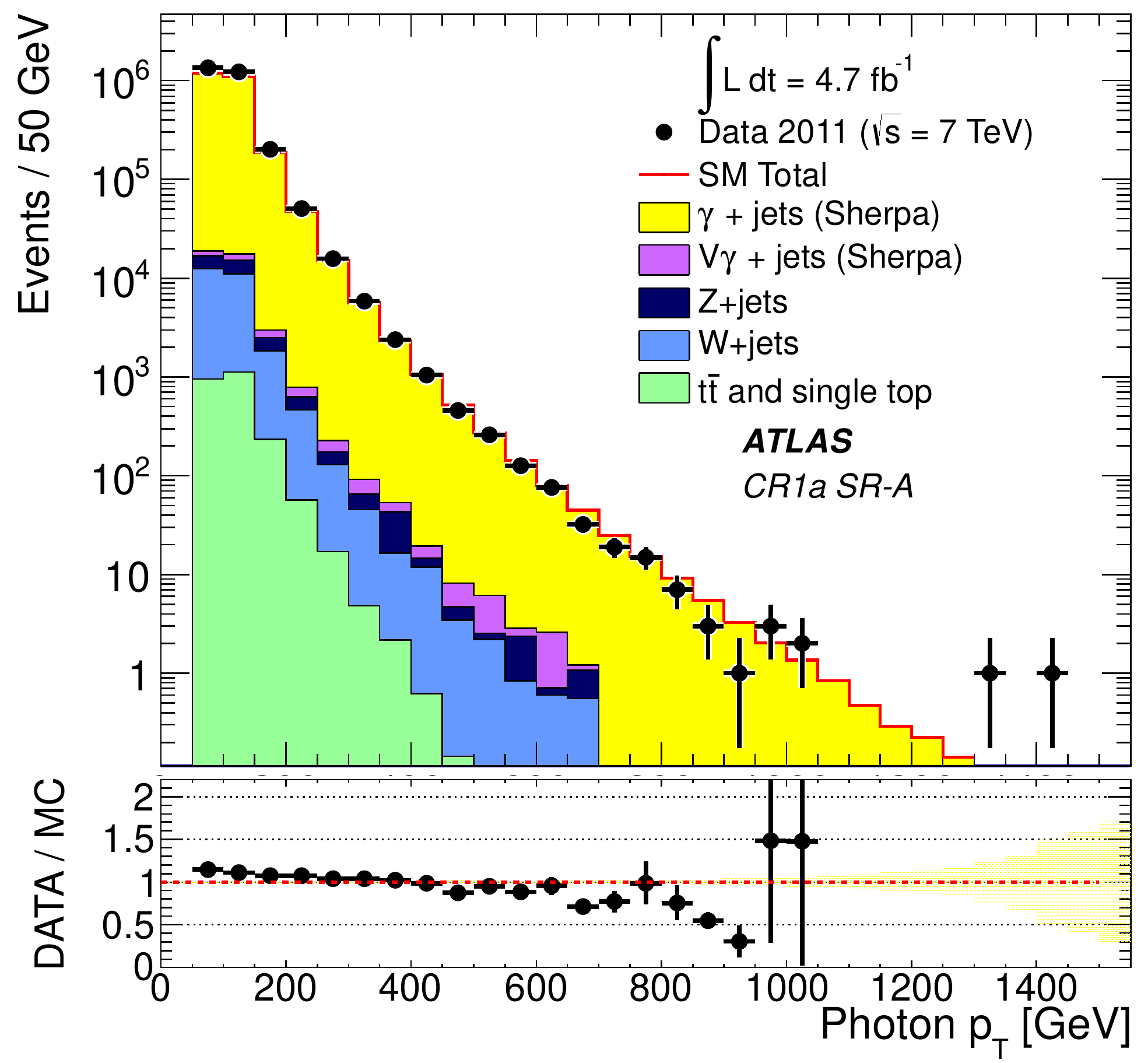}}
\subfigure[]{\label{fig:zfg:crD} \includegraphics[width=0.463\textwidth]{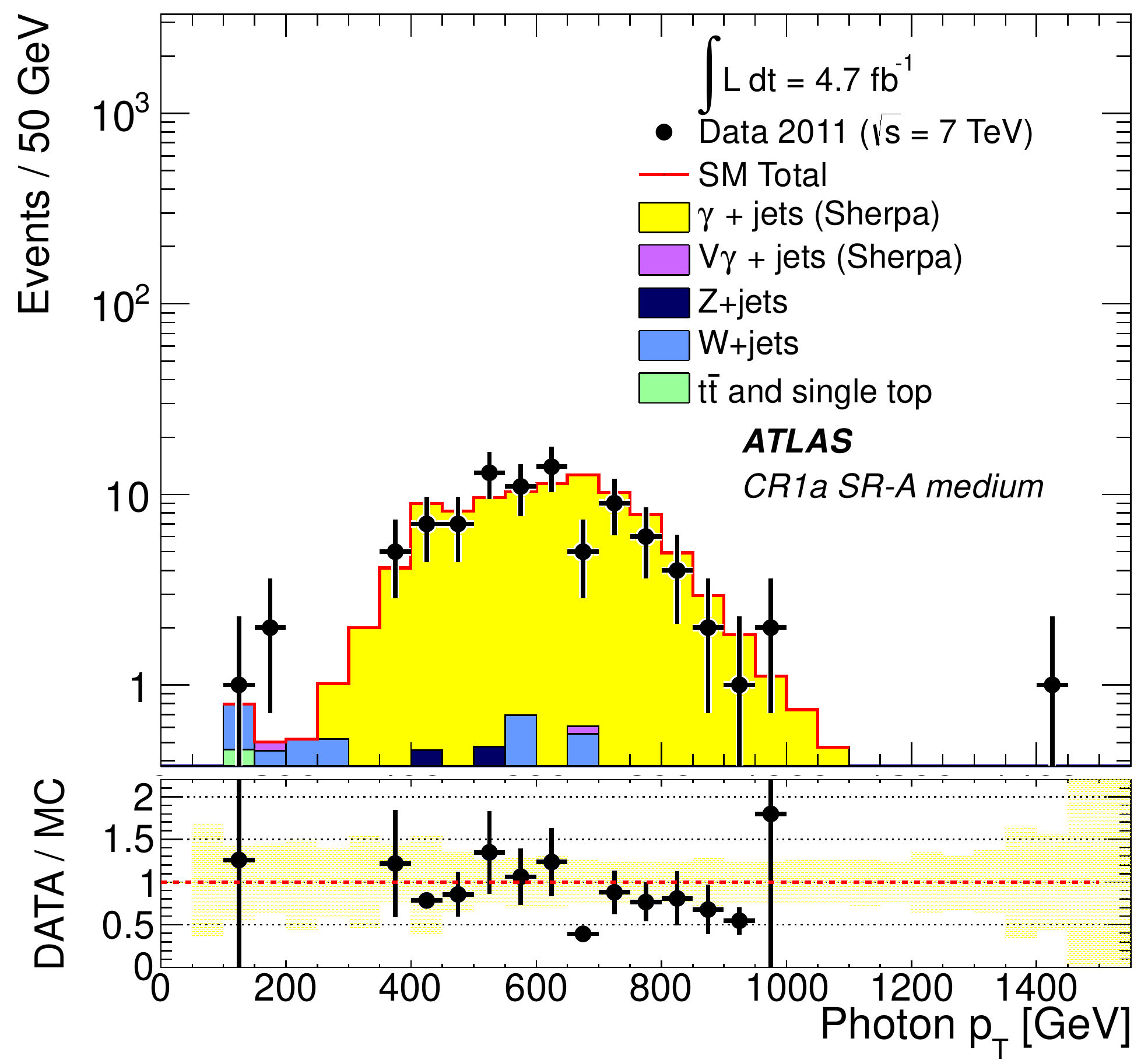}} \\ 
\caption{Leading photon $\pt$ distribution from data and MC simulation (a) directly after the photon 
selection and (b) in CR1a for SR-A medium that requires $\meff>1400$ GeV. The bottom panel shows the ratio of data to MC expectation, with the light (yellow) band indicating the uncertainty.
\label{fig:zfg:cr}}
\end{center}
\end{figure*}

Photon events are selected
by requiring at least one isolated photon passing the photon identification criteria discussed above. 
The photon trigger has an efficiency close to 100\% for selected events with a photon $\pt \geq85$~GeV. The photons are required to 
lie within the fiducial region $|\eta| < 1.37$ and $1.52 \leq |\eta| < 2.37$. After this 
first photon event selection a total of 2.8M photon candidates are obtained from the complete dataset, with an estimated purity $>$~95\%. Figure~\ref{fig:zfg:cr}(a)  shows the leading photon $\pt$ distribution for events passing the first photon selection.

In the second selection step, the SR selection criteria from Table~\ref{tab:srdefs} are applied to the photon sample. 
In order to prevent the reconstructed photon in the event from also being reconstructed as a jet, 
jets within $\Delta R = 0.2$ of the 
photon are removed. 
The photon $\pT$ is added to the $\met$ vectorial sum when applying the SR selections, using the appropriate calibration for the electromagnetic character of the photon shower.

The numbers of photon candidates which are selected by the CR1a criteria for channels A--E are 
presented in Table~\ref{tab:zfg:res} together with the numbers expected 
from MC simulation. Figure~\ref{fig:zfg:cr}(b) shows the leading photon $\pt$ distribution for events 
in CR1a for SR-A medium, that requires $\meff > 1400$ GeV. Good agreement is seen between the data and the MC simulation.

These numbers of photons are corrected for experimental effects as described in Ref.~\cite{ATLAS:2012ar} before being used to estimate the TFs. The following effects are considered. The combined identification and reconstruction efficiency is estimated to be 86\%, with an uncertainty 
of less than 1\%. The identification inefficiency is dominated by the tight photon identification 
requirements and decreases with increasing photon $\pT$. A further uncertainty of 5\% is included to account for differences in efficiency of the photon isolation criteria in different event samples.
Backgrounds from multi-jet processes and $W$+jets events where an electron from the $W$ decay is misidentified as a photon are each estimated to be $\sim$1\% for $p_{\rm T}^{\gamma} > 200$~GeV.  Therefore the background is neglected, but an uncertainty of 5\% is assigned.

\begin{table*}
\begin{center}
\begin{tabular}{| l | c | c | c | c | c | c |}
\hline
SR & Minimum \meff{}  & $\gamma$ CR1a data & $\gamma$ CR1a MC  & Est. $Z_{\nu \nu}$ SR ($\gamma$) data \\
      &        GeV         &                                          & {\small \sherpa/\alpgen} &                                                                          \\
\hline
 \multirow{2}{*}{A} 
     & 1400          & 90                                     & 96 / 93.4                         & 32.0 $\pm$ 3.4     $\pm$ 5.6    \\
     & 1900          & 9                                       & 9.42 / 9.33                      & 3.2   $\pm$ 1.1     $\pm$ 0.6     \\
  \hline
 A$'$
     & 1200          & 170                                  & 176 / 180                        & 62    $\pm$ 5         $\pm$ 11        \\
  \hline
 B
     & 1900          & 5                                       & 6.21 / 6.31                      & 1.9  $ \pm$ 0.8     $\pm$ 0.4         \\
 \hline
 \multirow{3}{*}{C}
    & 900             & 223                                  & 219 / 197                         & 64   $\pm$ 4      $\pm$ 11          \\
    & 1200           & 48                                    & 55.8 / 44.5                       & 15 $\pm$ 2        $\pm$ 3         \\
    & 1500           & 6                                      & 14.4 / 11.1                       & 1.9  $\pm$ 0.8    $\pm$ 0.4       \\
  \hline
 D
    & 1500           & 3                                      & 10.9 / 6.98                       & 0.86  $\pm$ 0.50    $\pm$ 0.24  \\
   \hline
 \multirow{3}{*}{E} 
   & 900              & 77                                    & 71.5 / 47.4                       & 20 $\pm$ 2      $\pm$ 5                \\
   & 1200            & 26                                   & 15.3 / 13.9                        & 7.7 $\pm$ 1.5  $\pm$ 1.9        \\
   & 1400            & 11                                   & 8.71 / 6.11                        & 3.4 $\pm$ 1.0  $\pm$ 1.0      \\
 \hline
\end{tabular}
\caption{Numbers of photon events observed in the data and expected from the \sherpa{} and \alpgen{} MC simulations in CR1a for each SR, as well as the resulting estimated numbers 
of $Z (\rightarrow\nu\bar{\nu})$ events in the SRs, with statistical and systematic uncertainties. 
}
\label{tab:zfg:res}
\end{center}
\end{table*}

The number of photon events selected by the CR1a criteria is used to estimate 
the expected number of $Z (\rightarrow \nu\bar{\nu})$ events in the corresponding SR using

\begin{eqnarray}
N^{Z (\rightarrow \nu\bar{\nu})}(p_{\rm T}) & = & N^{\gamma}(p_{\rm T}) \cdot \Biggl[
\frac{(1 - f_{\rm bkg})}
{\varepsilon^{\gamma}(p_{\rm T}) \cdot A^{\gamma}(p_{\rm T})} \cr
& & \phantom{}\cdot R_{Z/\gamma}(p_{\rm T}) \cdot Br(Z \rightarrow \nu\bar{\nu})  \Biggr] .
\label{eq:gamma}
\end{eqnarray}

Here $N^{\gamma}(p_{\rm T})$ represents the number of photon candidate events passing the CR1a selections, binned in $\pt$ as in Fig.~\ref{fig:zfg:cr}(b), $ f_{\rm bkg}$ the fraction of fake photons in the control region, $\varepsilon^{\gamma}(p_{\rm T})$ the efficiency for selecting the photons and $A^{\gamma}(p_{\rm T})$ the photon acceptance. 
The cross section ratio $R_{Z/\gamma}(p_{\rm T})$ is determined from MC simulation.
The uncertainties related to the cross section ratio have been studied using the two 
MC programs {\tt PYTHIA8}~\cite{Sjostrand:2007gs} and {\tt GAMBOS} (an adaptation of the {\tt VECBOS} program~\cite{Ask:2011xf, Berends:1990ax}) and many of the theoretical uncertainties, such 
as the choice of scales and parton distribution functions, are found to 
cancel in the ratio, to a large extent~\cite{Ask:2011xf}. It has, however, been shown that the ratio retains slight sensitivity 
to the jet selection and that multi-parton matrix elements must be used to 
describe correctly all the relevant amplitudes. The final uncertainties on 
$R_{Z/\gamma}(p_{\rm T})$ should therefore be small, but a conservative uncertainty of 25\% is assigned. Additional systematic uncertainties, common to several parts of the analysis, are discussed in Section~\ref{sec:systematics}.

 The transfer factors between the CR1a regions and their associated signal regions are obtained by averaging the correction term in the square brackets of Eq.~\ref{eq:gamma} over the measured $\pt$ distribution of selected photon candidates, and are given in Table~\ref{tab:TFallWithErrors}.

\subsection{\texorpdfstring{$Z$}{Z}+jets estimate using a \texorpdfstring{$Z(\to \ell\ell)$}{Z to ll} + jets control region} \label{zjet}

The irreducible background from $Z (\rightarrow\nu\bar{\nu})$+jets can also be estimated independently using the observed leptonic $Z$ decays. The CR1b control regions are defined by requiring two opposite-sign electrons or muons with $p_{\rm T}>20$~\GeV. In addition, the $\pt$ of the leading electron is required to be above 25~\GeV{} to protect against trigger turn-on effects.  The di-lepton invariant mass must lie in the range $66$~\GeV~$< m(\ell\ell) < 116$~\GeV. The \met\ variable in the SR selection is emulated with the vectorial sum of the reconstructed $Z$ boson momentum vector and the measured $\ourvecptmiss$.  The SR jet and \met\ requirements are applied, without selections on $\ourdeltaphifull$ or \met/\meff . These changes are made to increase the acceptance, since the precision of the method is limited by the rate of di-lepton events.

In order to calculate the transfer factors, \alpgen~is used to estimate the number of $Z$+jets events in each SR and the number of $Z(\rightarrow \ell\ell)$+jets events in each corresponding CR1b control region. 
The uncertainties arise from the number of MC simulation events, the jet energy scale and resolution, the electron and muon energy resolutions, the electron and muon selection efficiencies, the electron trigger efficiency, the electron energy scale, energy included in calorimeter clusters that is not associated with physics objects, the input PDFs, the modeling of pile-up in the simulation, and the luminosity. 

The transfer factors themselves are listed in Table~\ref{tab:TFallWithErrors} and take into account the contribution from leptonic $Z(\to\tau\tau)$+jets events in CR1b. The estimated numbers of $Z$+jets events obtained using this technique are consistent with those estimated using $\gamma$+jets events observed in CR1a.

\subsection{Multi-jet background estimation}
\label{subsec:qcd}
The probability for multi-jet events to pass any of the SR selection cuts used in this analysis is, by design, very small. However, the large cross section for this process could potentially compensate for the low acceptance and hence lead to significant SR contamination.  These two effects also limit the applicability of conventional MC simulation techniques; firstly because very large MC data samples are required and secondly because accurate modeling of the acceptance requires exceptionally detailed understanding of the performance of every component of the calorimeters. For this reason a data-driven method is used to estimate the multi-jet background in the SRs. The method makes use of high-statistics samples of well-measured data multi-jet events to minimize statistical uncertainties. In order to determine the acceptance of the SRs for poorly-measured multi-jet events, the jets in these events are convoluted with a function modeling the response of the calorimeters. This response function is based upon the results of MC simulations but is modified in such a way as to give good agreement between multi-jet estimates and data in two additional dedicated analyses. This procedure minimizes the susceptibility of the multi-jet background estimates in the main analysis to systematic uncertainties arising from the Monte Carlo modeling of the initial response function. 

The jet response function quantifies the probability of fluctuation of the measured \pt{} of jets and takes into account both the effects
of jet mismeasurement and contributions from neutrinos and muons in jets from
heavy flavor decays. This function is convoluted with the
four-vectors of jets in low-\met{} multi-jet data events, generating higher \met{}
events. These are referred to as `pseudo-data' and are used to provide a minimally MC simulation dependent
estimate of multi-jet distributions, including the distribution of $\ourdeltaphifull$ for high $\meff$ events. These distributions can be used to determine the transfer factors from the low $\ourdeltaphifull$ multi-jet control regions CR2 to the higher $\ourdeltaphifull$ signal regions.

The method, referred to as the `jet smearing method' below, proceeds in four steps:
\renewcommand{\labelenumi}{(\arabic{enumi})}
\begin{enumerate}
  \item Selection of low-\met{} {\em seed events} in the data. The jets in these events are well measured. These events are used in steps (3) and (4).
  \item As a starting point
   the response function is determined in MC simulated data by comparing generator-level jet energy to reconstructed detector-level jet energy.  
  \item Jets in the seed events are convoluted with the response function to generate pseudo-data events. The consistency between pseudo-data and experimental data in two analyses (see below) is then determined. The response function is modified and the convolution repeated until good agreement is obtained.
  \item Jets in the seed events are convoluted with the resulting data-constrained response function to obtain a final sample of pseudo-data events. This sample is used to estimate the distributions of
  variables defining the control and signal regions used in the main analysis.
\end{enumerate}
\noindent

Seed events are triggered using single-jet triggers and offline thresholds of 50, 100, 130, 165, 200, 260 and 335~GeV are then applied. To ensure
that the events contain only well-measured jets, the \met{} significance (defined as $\met/\sqrt{E_{\rm T}^{\rm sum}}$ where $E_{\rm T}^{\rm sum}$ is the scalar sum of the transverse energy measured in the calorimeters) is required to be $<0.6$~\GeV$^{1/2}$. 

The response function is initially estimated from MC simulation by matching `truth'
jets reconstructed from generator-level particles to detector-level jets with $\Delta R<0.1$ in multi-jet samples. The
four-momenta of any generator-level neutrinos in the truth jet cone are
added to the four-momentum of the truth jet. Truth
jets are isolated from other truth jets by $\Delta R>0.6$. The
response is the ratio of the reconstructed detector-level to generator-level jet transverse energy.

A `smeared' event is generated by multiplying each jet four-momentum in
a seed event by a random number drawn from the response function. The
smeared event \met{} is computed using the smeared transverse momenta of the jets.
The response function measured using MC simulation is modified using additional Gaussian smearing to widen the jet response, and a correction is applied to the low-side response tail to adjust its shape. These corrections improve the agreement with the data in step (3).
   
Two dedicated analyses are used to constrain the shape of the jet
response function in step (3). The first uses the
\pt{} asymmetry of di-jet events. Events with two jets with
$\vert\eta\vert<2.8$ and \pt{} $>70$, 50~\GeV{} are selected, where there
are no additional jets with $\vert\eta\vert<2.8$ and $\pt$~$>$~40~GeV. Events
are vetoed if they contain any jet with $\pt$~$>$~20~GeV and $\eta>2.8$. The
\pt{} asymmetry is given by
\begin{equation} \label{eq:smearing/ptAssym}
A(p_{\mathrm{T},1},p_{\mathrm{T},2})=\frac{p_{\mathrm{T},1}-p_{\mathrm{T},2}}{p_{\mathrm{T},1}+p_{\mathrm{T},2}},
\end{equation}
where the indices correspond to the jet \pt{} ordering. This distribution is sensitive to the Gaussian response of the
jets and to any non-Gaussian tails. 
A fit of pseudo-data to the collision data asymmetry distribution is used to adjust the response function generating the pseudo-data.

\begin{figure*}[htb]
  \begin{center}
  \subfigure[]{
    \includegraphics[width=0.50\figwidth]{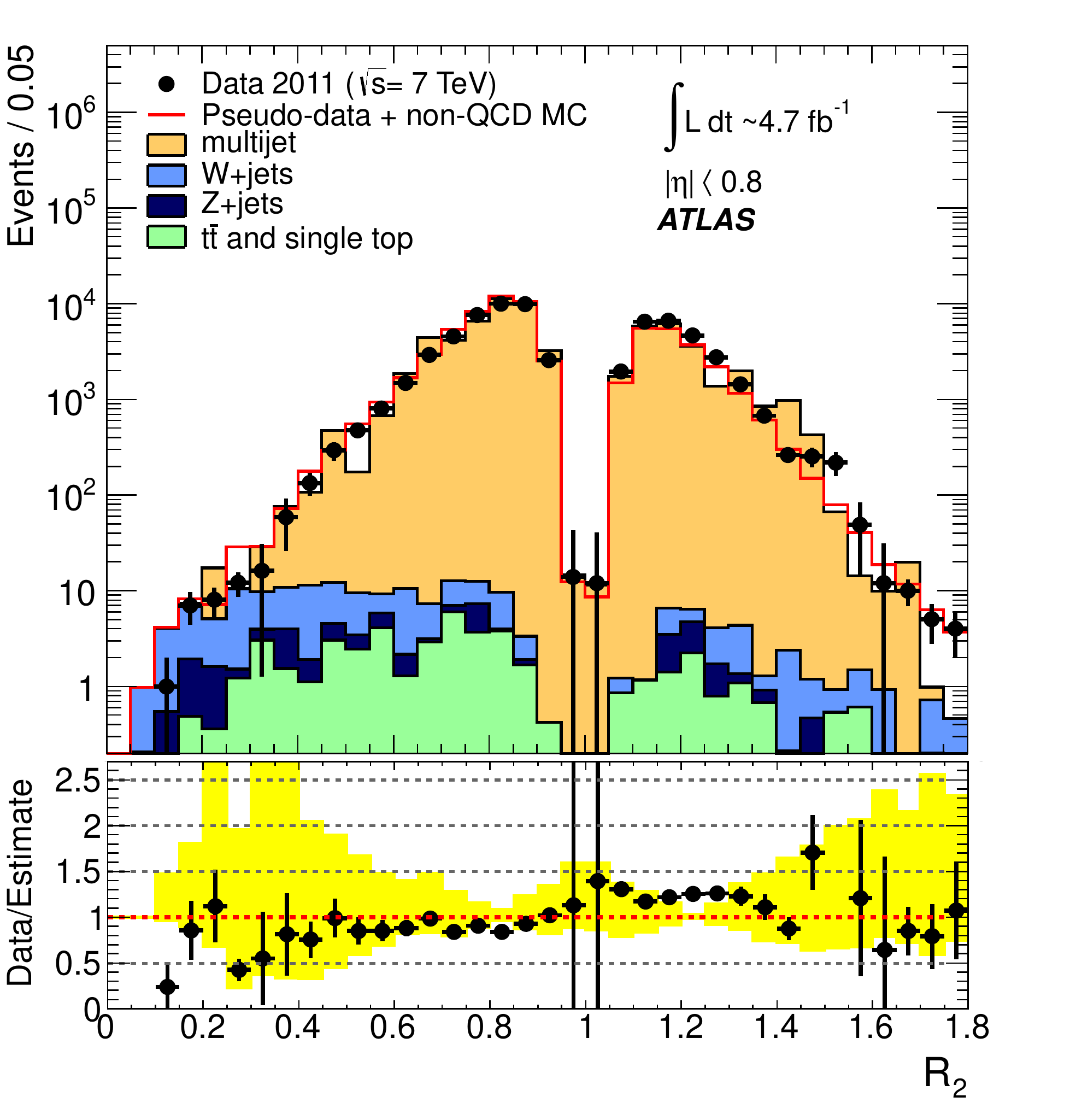}
    }
  \subfigure[]{
    \includegraphics[width=0.50\figwidth]{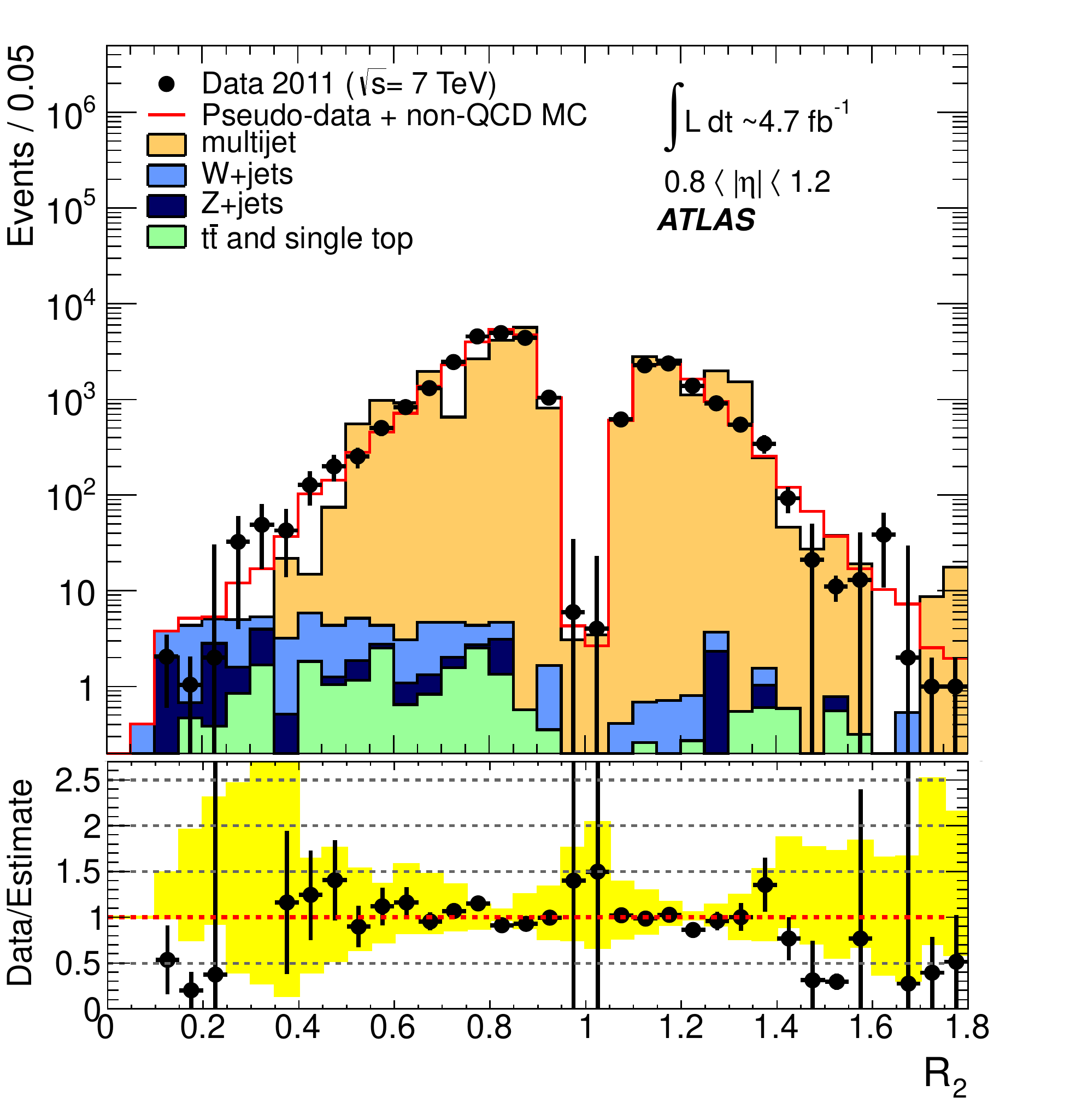}
    }
  \subfigure[]{
    \includegraphics[width=0.50\figwidth]{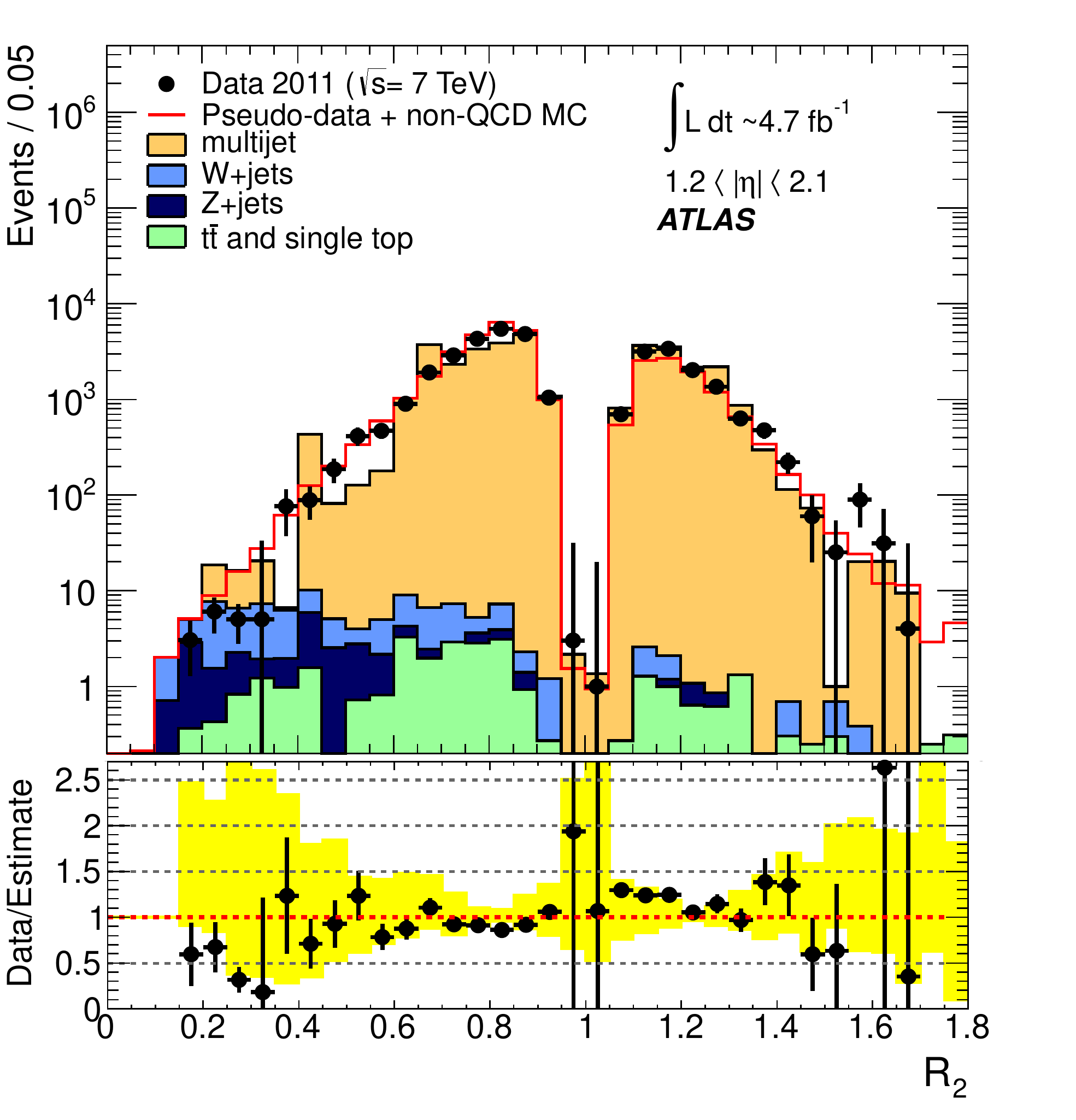}
    }
  \subfigure[]{
    \includegraphics[width=0.50\figwidth]{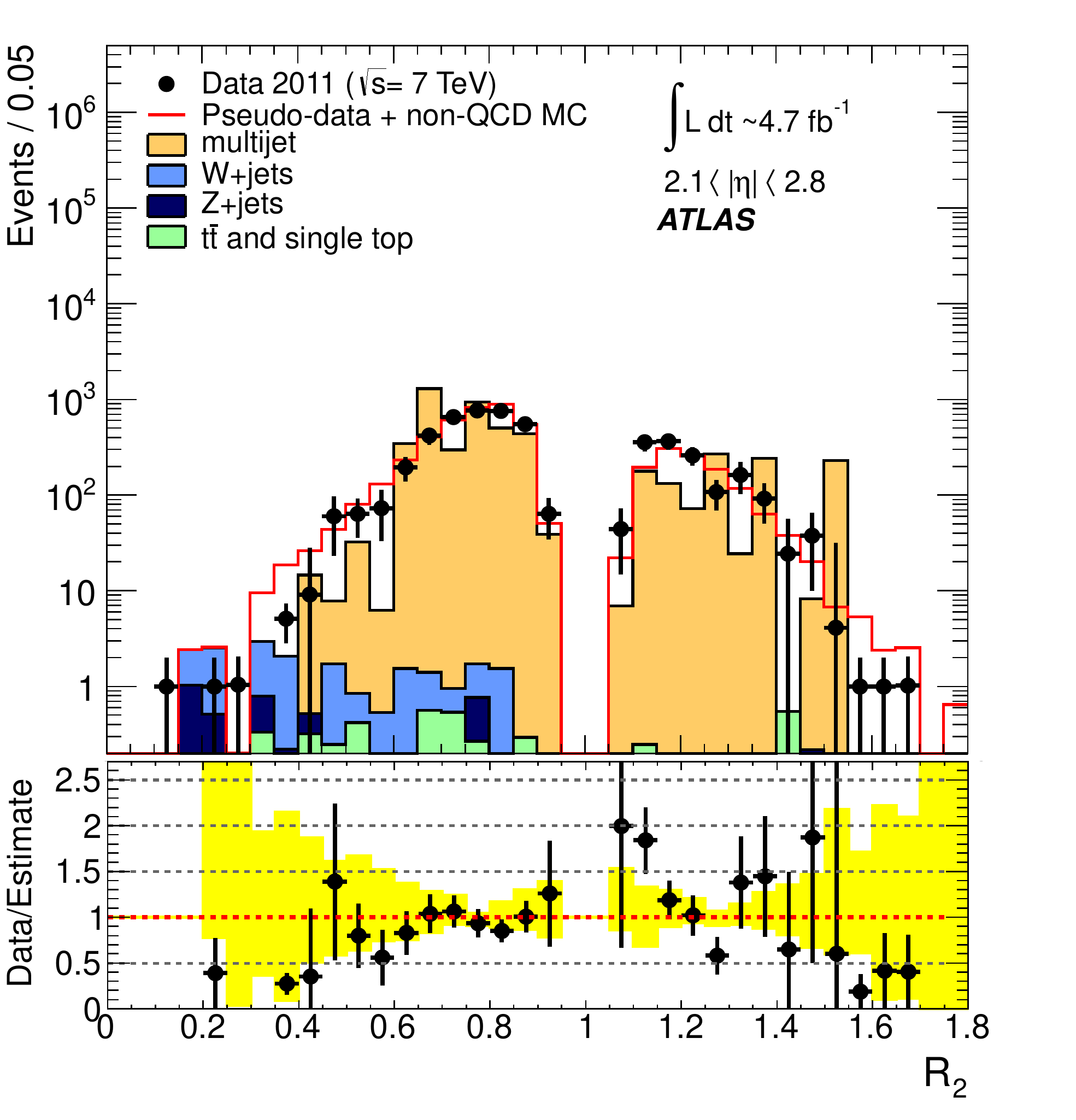}
    }

 \end{center}
\caption[Mercedes plots]{ \label{fig:smearing/merc_eta}
Distributions of $R_2$ in four bins (a-d) of $|\eta|$ of the poorly reconstructed jet, for estimated true jet $\pT$, defined as $|\vec{p}_{\mathrm{T}}^J + \ourvecptmiss|$, greater than $100\
\mathrm{GeV}$. The black points represent collision data while the open medium (red) histogram represents the combined prediction. The jet smearing method described in the text is used to estimate the multi-jet contribution (referred to in the plots as ``pseudo-data'') while MC simulation predictions are used for the other background components.
The lower panels show the fractional deviation of the data from the prediction (black points), with the light (yellow) bands showing the multi-jet uncertainty combined with the MC simulation statistical uncertainty on the non-multi-jet estimate.}  
\end{figure*}

A second analysis
studies the $R_2$ distribution of $\geq 3$-jet events where
topological selections ensure that one jet is unambiguously associated in $\phi$ with
the \met{} in the event. The response of the detector to this jet is then given approximately by the quantity $R_2$ defined by
\begin{equation} \label{eq:tailRes}
  R_2 \equiv \frac{\vec{p}_{\mathrm{T}}^J \cdot (\vec{p}_{\mathrm{T}}^J + \ourvecptmiss)}
                  {|\vec{p}_{\mathrm{T}}^J + \ourvecptmiss|^2},
\end{equation}
where $\vec{p}_{\mathrm{T}}^J$ is understood to be the \emph{reconstructed} \pt{}
of the jet associated with the \met{}. This distribution is sensitive to the tails of the response function from mismeasured jets. When the $\pt$~of the jet is under-measured, $\ourvecptmiss$ lies parallel to $\vec{p}_{\mathrm{T}}^J$ and hence $R_2<1$. Conversely, when the $\pt$~of the jet is over-measured, $\ourvecptmiss$ lies anti-parallel to $\vec{p}_{\mathrm{T}}^J$ and hence $R_2>1$. Fits are performed in \pt{} and $\eta$ bins in order to constrain the parameters describing the low-side response function tail, which affects primarily the region with $R_2 \ll 1$. 

The $R_2$ distribution provides a sensitive test of the response function and hence of the background estimate in different regions of the detector, such as the transition between the barrel and end-cap calorimeters, where the energy resolution is degraded by the presence of dead material. 
The data are divided into four regions according to the $\eta$ of the poorly reconstructed jet associated with the \met, shown in Fig.~\ref{fig:smearing/merc_eta}. The
estimates agree well, with the data indicating that non-Gaussian
fluctuations are not strongly $\eta$ dependent. Given the good agreement observed between the
data and estimates, no uncertainty is associated with the $\eta$
dependence of the response. Following this procedure, a good estimate of the jet response function, including non-Gaussian tails, is obtained.

\begin{figure*}
  \begin{center}
  \subfigure[]{
    \includegraphics[width=0.43\textwidth]{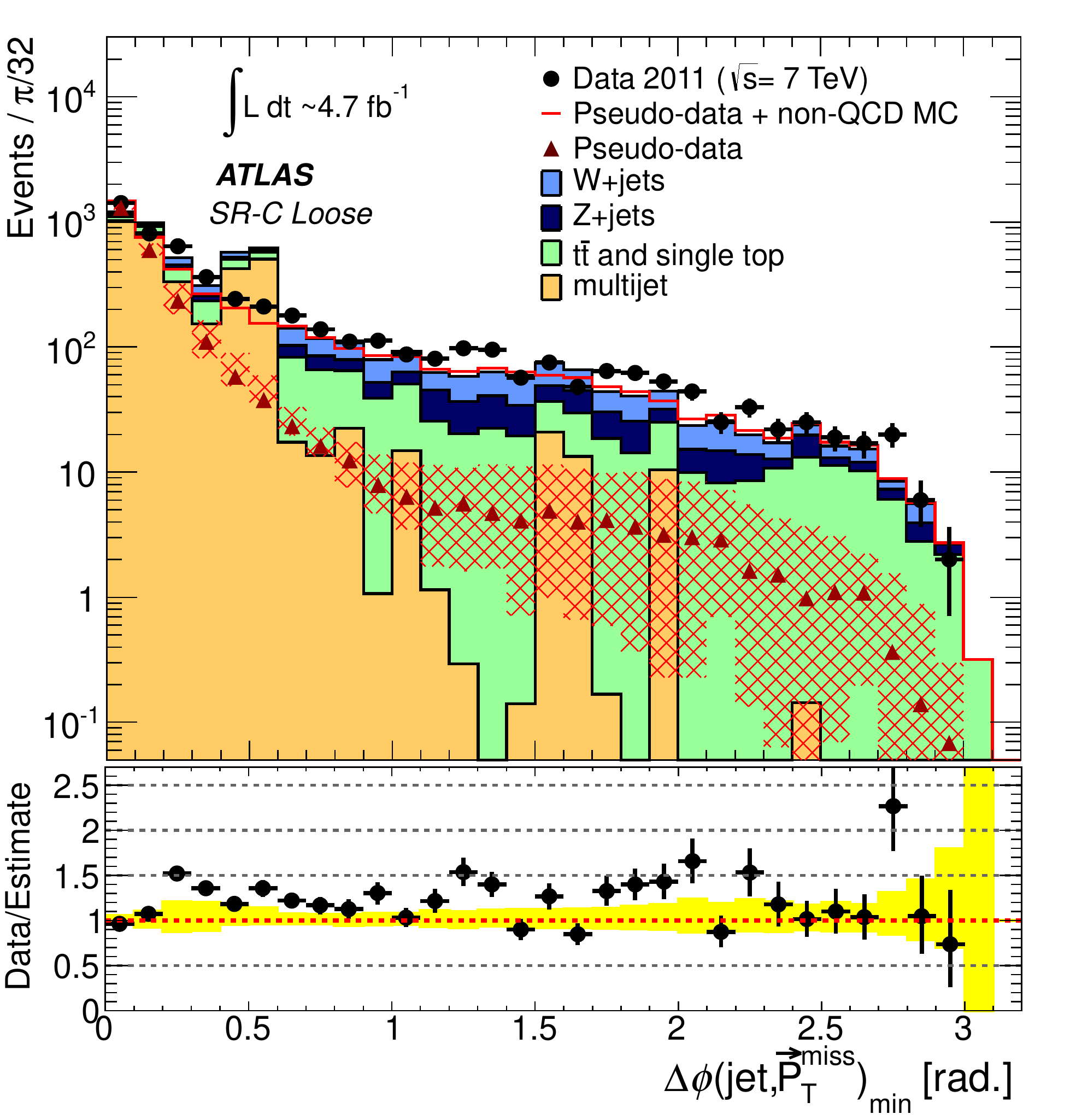}
    }
  \subfigure[]{
    \includegraphics[width=0.43\textwidth]{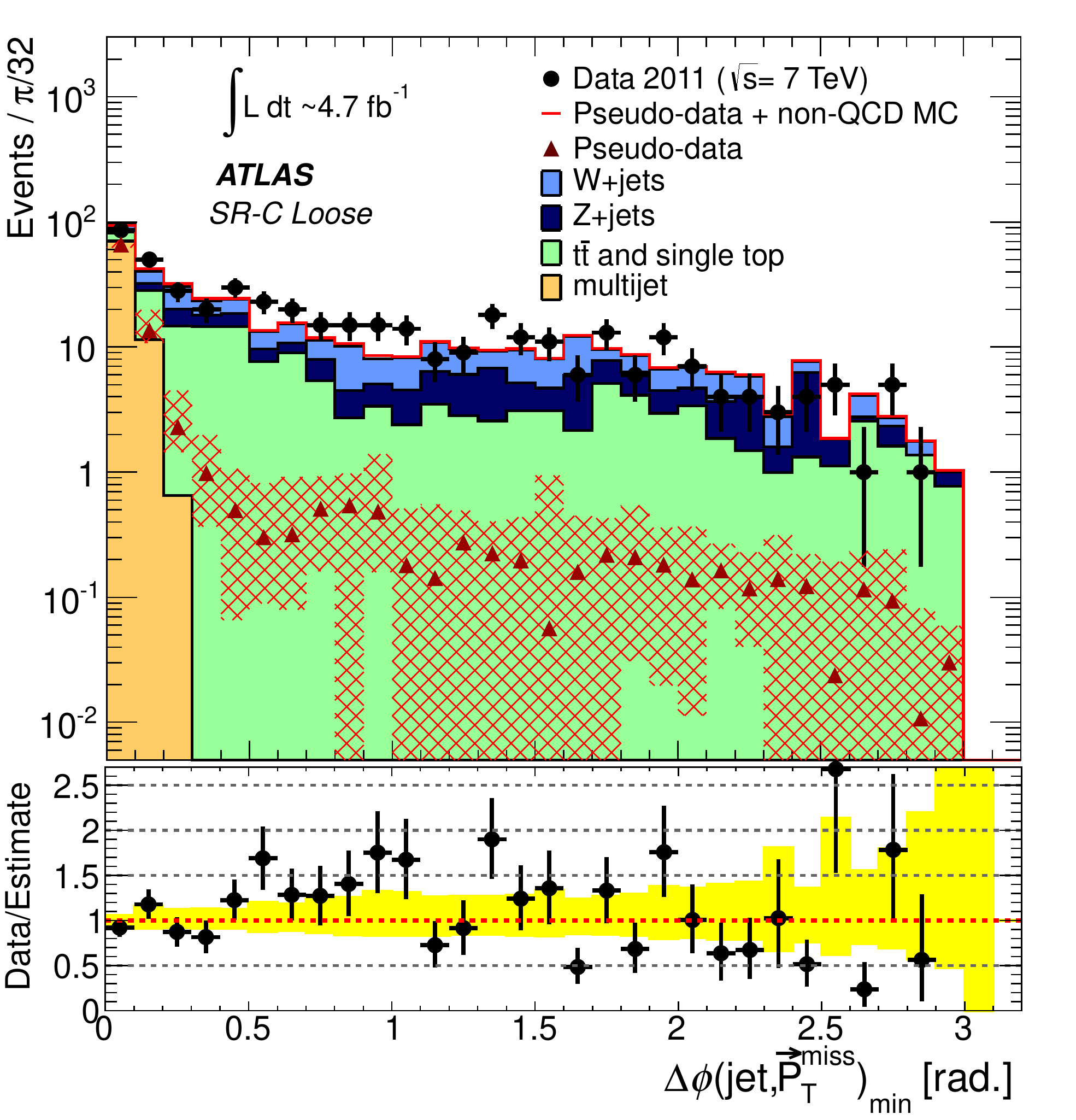}
    }
\end{center}
\vspace{-6mm}
\caption{ Comparison of observed and predicted
distributions of $\ourdeltaphifull$ for the leading three jets ($\ourdeltaphifull$ $(i=\{1,2,3\})$), (a) after all selections except for those on $\ourdeltaphifull$, $\meff$
and $\met/\meff$ and (b) after all selections except for that on $\ourdeltaphifull$
for signal region C loose. The histograms show the MC simulation estimates of each background component. The medium (maroon) triangles show the multi-jet estimates from the jet smearing technique, normalized in the regions with $\ourdeltaphifull$ $(i=\{1,2,3\})<0.2$ radians, which replaces the multi-jet MC simulation estimate (denoted with a histogram) in the main analysis. The hatched region denotes the total uncertainty on the multi-jet estimate including statistical uncertainties from the seed event sample and the smearing procedure, systematic uncertainties in the jet response function, and bias in the seed event selection.
The lower panels show the fractional deviation of the data from the prediction (black points), with the light (yellow) bands showing the multi-jet uncertainty combined with the MC simulation statistical uncertainty on the non-multi-jet estimate.
}
\label{PredictedPlots}
\end{figure*}

In order to illustrate the technique, Fig.~\ref{PredictedPlots} shows comparisons between SM MC simulation
predictions, data and the jet smearing estimate for distributions of $\ourdeltaphifull$ calculated with just the leading three jets. The figure makes use of the earlier stages of the event
selections for SR-C loose and its associated multi-jet control region. The final event selections used in the analysis impose further requirements on $\ourdeltaphifull$ for additional jets with $\pt > 40$~\GeV{} (see Table~\ref{tab:srdefs}). Good agreement is seen in Fig.~\ref{PredictedPlots} both between the data and MC simulation, and between the data
and the smearing estimate. 

In order to check that the above method is robust against changes in pile-up conditions, which changed significantly during data-taking, the method was repeated with the data divided into sub-samples corresponding to four time periods representative of different pile-up regimes. No significant dependence upon the level of pile-up was found.

The resulting multi-jet transfer factors between CR2 and SR for the signal regions are shown in Table~\ref{tab:TFallWithErrors}.

\subsection{\texorpdfstring{$W(\to\ell\nu)$}{W to lnu}+jets and \texorpdfstring{$t\bar{t}$}{ttbar} background estimation}
\label{subsec:wtop}
The lepton veto applied to the signal events aims to suppress SM events with an isolated lepton. However, 
such a veto does not reject all \ttbar~and $W$+jets events, particularly when their decay products 
involve a lepton which is out of acceptance, or not reconstructed, or when the 
lepton is a hadronically-decaying $\tau$. 

To estimate the contributions from $W$+jet and top quark backgrounds in the signal regions,
two CRs are defined for each SR, one with a $b$-jet veto (CR3 -- enriched in $W$+jets events) and one with a $b$-tag
requirement (CR4 -- enriched in \ttbar~events) as defined in Table~\ref{tab:crdefs}. With the exception of the $b$-jet requirement/veto the selections for CR3 and CR4 are identical and hence the two samples are fully anti-correlated. Both of these CRs require exactly one `signal' electron or muon satisfying tighter selection criteria, whose transverse mass, formed with the \met , lies between 30~\GeV{} and 100~\GeV. The lepton is then modeled as an additional jet, as it would be if it had entered the signal regions. The $\ourdeltaphifull$ and $\met/\meff$ criteria which are applied in the corresponding signal regions are not applied to
the CRs, in order to increase the CR sample sizes. 

In the electron channel, the modeling of the lepton as a jet is 
physically accurate, as the reconstruction will interpret misidentified electrons in this way. In 
the muon case, a missed muon will contribute additional missing transverse momentum, 
rather than an extra jet (although a small fraction of its energy may well be deposited in the
calorimeters). When the lepton is a hadronically-decaying tau, the behavior lies between these two extremes, with the hadrons being seen as jet activity and the $\tau$-neutrino as missing momentum. In order to be consistent between the electron and muon channels, and to use one high-statistics control region each for top quark and $W$ events, the choice is made to model 
all leptons as jets. This is justified by the fact that the majority of the background comes from hadronic $\tau$-decay events, for which the behavior of the lepton is more jet-like than \met-like. It should be noted that this choice does not bias the background estimate because identical procedures are applied to data and to MC simulation events used to construct the transfer factors. The procedure has been validated with two alternative choices, in which the lepton is modeled either as missing transverse momentum or as a $\tau$ decay. 

The transfer factors are calculated using MC simulation. Several corrections are applied to MC simulation events:
\begin{itemize}
\item Each event in the CR is weighted by the ratio of the lepton identification efficiency
in data to that in simulation. Similarly, the 
numbers in the signal region are weighted by a corresponding {\it in}efficiency scale factor. This weighting is 
performed on an event-by-event basis, based on the simulated lepton's transverse momentum and pseudorapidity.
\item A similar scale factor is applied for the $b$-tagging efficiency (CR4) and fake rate (CR3), which 
differ between data and simulation~\cite{ATLAS-CONF-2011-102,ATLAS-CONF-2012-043}. This is also performed as an event-by-event weighting.
\item The leptons are smeared such that their energy resolution reflects that measured in data.
\end{itemize}

Various sources of systematic uncertainty on the transfer factors have been considered. For the leptons, the identification efficiency, energy resolution and trigger efficiency are considered. The $b$-tagging efficiency and fake rate, jet energy scale and jet energy resolution (for both $b$-quark and light jets separately), are considered, together with the effect of pile-up, of calorimeter electronics failures and of calorimeter energy deposits not associated with physics objects.
The fake lepton background is found to be negligible in both CR3 and CR4. 

The TFs between CR3, CR4, and the signal 
regions are given in Table~\ref{tab:TFallWithErrors}. Similar TFs are also computed for each channel between CR3, CR4
and the multi-jet control region CR2, where $W$+jets and $t\bar{t}$ events can contribute significantly.

\subsection{Estimated transfer factors}
\label{transfersummary}
The transfer factors estimated using the methods described above are summarized in Table~\ref{tab:TFallWithErrors} for each CR. These values, and those between the various CRs, together with the observed event counts in each SR and CR form the inputs to the likelihood fit described in Section~\ref{sec:stats}.

\begin{table*}
\begin{center}\renewcommand\arraystretch{1.4}
\begin{tabular}{|c|c|c|c|c|c|}
\hline
\multirow{2}{*}{Signal region}     &\multicolumn{5}{|c|}{Control region / Process} \\ 
\cline{2-6}
& CR1a / $Z$+jets  & CR1b / $Z$+jets  & CR2 / Multi-jets   & CR3 / $W$+jets & CR4 / $t\bar{t}$+ single top  \\
\hline
\hline
SR-C loose & $ 0.32 ^{+ 0.08 }_{- 0.09 }$
& $ 2.9 ^{+ 0.7 }_{- 0.4 }$
& $ 0.016 ^{+ 0.012 }_{- 0.012 }$
& $ 0.36 ^{+ 0.04 }_{- 0.05 }$
& $ 0.52 ^{+ 0.08 }_{- 0.08 }$
\\
SR-E loose & $ 0.27 ^{+ 0.08 }_{- 0.08 }$
& $ 6.5 ^{+ 5.0 }_{- 3.0 }$
& $ 0.05 ^{+ 0.04 }_{- 0.04 }$
& $ 0.74 ^{+ 0.12 }_{- 0.13 }$
& $ 0.92 ^{+ 0.18 }_{- 0.19 }$
\\
SR-A medium & $ 0.36 ^{+ 0.10 }_{- 0.10 }$
& $ 2.5 ^{+ 0.7 }_{- 1.0 }$
& $ 0.032 ^{+ 0.019 }_{- 0.019 }$
& $ 0.31 ^{+ 0.05 }_{- 0.05 }$
& $ 0.34 ^{+ 0.22 }_{- 0.22 }$
\\
SR-A' medium & $ 0.39 ^{+ 0.10 }_{- 0.10 }$
& $ 2.2 ^{+ 0.5 }_{- 0.6 }$
& $ 0.10 ^{+ 0.06 }_{- 0.06 }$
& $ 0.19 ^{+ 0.03 }_{- 0.02 }$
& $ 0.23 ^{+ 0.06 }_{- 0.07 }$
\\
SR-C medium & $ 0.34 ^{+ 0.09 }_{- 0.10 }$
& $ 2.9 ^{+ 1.8 }_{- 1.0 }$
& $ 0.003 ^{+ 0.005 }_{- 0.001 }$
& $ 0.20 ^{+ 0.06 }_{- 0.05 }$
& $ 0.30 ^{+ 0.10 }_{- 0.10 }$
\\
SR-E medium & $ 0.32 ^{+ 0.10 }_{- 0.10 }$
& $ 5.0 ^{+ 9.0 }_{- 3.0 }$
& $ 0.038 ^{+ 0.031 }_{- 0.031 }$
& $ 0.39 ^{+ 0.10 }_{- 0.10 }$
& $ 0.62 ^{+ 0.17 }_{- 0.19 }$
\\
SR-A tight & $ 0.30 ^{+ 0.08 }_{- 0.08 }$
& $ 5.3 ^{+ 4.1 }_{- 3.7 }$
& $ 0.009 ^{+ 0.009 }_{- 0.009 }$
& $ 0.25 ^{+ 0.09 }_{- 0.10 }$
& $ 0.01 ^{+ 0.02 }_{- 0.02 }$
\\
SR-B tight & $ 0.38 ^{+ 0.10 }_{- 0.10 }$
& $ 4.2 ^{+ 3.7 }_{- 5.3 }$
& $ 0.011 ^{+ 0.008 }_{- 0.008 }$
& $ 0.14 ^{+ 0.07 }_{- 0.08 }$
& $ 0.022 ^{+ 0.023 }_{- 0.024 }$
\\
SR-C tight & $ 0.32 ^{+ 0.09 }_{- 0.09 }$
& $ 1.8 ^{+ 1.4 }_{- 0.9 }$
& $ 0.0034 ^{+ 0.0044 }_{- 0.0025 }$
& $ 0.16 ^{+ 0.11 }_{- 0.11 }$
& $ 0.15 ^{+ 0.12 }_{- 0.13 }$
\\
SR-D tight & $ 0.29 ^{+ 0.10 }_{- 0.08 }$
& $ 2.1 ^{+ 4.8 }_{- 2.8 }$
& $ 0.02 ^{+ 0.01 }_{- 0.01 }$
& $ 0.26 ^{+ 0.10 }_{- 0.10 }$
& $ 0.20 ^{+ 0.15 }_{- 0.16 }$
\\
SR-E tight & $ 0.31 ^{+ 0.11 }_{- 0.10 }$
& $ 2.7 ^{+ 2.8 }_{- 4.5 }$
& $ 0.04 ^{+ 0.02 }_{- 0.02 }$
& $ 0.26 ^{+ 0.10 }_{- 0.08 }$
& $ 0.32 ^{+ 0.27 }_{- 0.25 }$
\\
  \hline
 \end{tabular}
\caption{\label{tab:TFallWithErrors} Summary of transfer factors from the main control regions of each background component in every stream. In CR4 for signal regions A tight and B tight the $\meff$ requirements were relaxed to 1500 GeV to increase the numbers of events in the CRs for the minor $t\bar{t}$ background.}
 \end{center}
 \end{table*}

\section{Systematic Uncertainties}\label{sec:systematics}

Systematic uncertainties arise through the use of the transfer factors relating observations in the control regions to background expectations in the signal regions, and from the modeling of the SUSY signal. For the transfer factors derived from MC simulation the primary common sources of systematic uncertainty are the jet energy scale (JES) calibration, jet energy resolution (JER), MC modeling and statistics, and the reconstruction performance in the presence of pile-up.
 
The JES uncertainty has been
measured from the complete 2010 dataset using the techniques
described in Ref.~\cite{atlas-jes-paper2011} and is around 4\%,
with a slight dependence upon $\ourpt$, $\eta$ and the proximity to adjacent jets.
The JER uncertainty is estimated using the methods discussed in Ref.~\cite{atlas-jes-paper2011}.
Additional contributions are added to both the JES and the JER uncertainties to take account of the effect of pile-up at the relatively high luminosity delivered by the LHC in the 2011 run. 
Both in-time pile-up arising from multiple collisions within the same bunch-crossing, and out-of-time pile-up, which arises from the detector response to neighboring bunch crossings, are taken into account.

The dominant modeling uncertainty in the MC simulation estimate of the numbers of events in the signal and control regions arises from the impact of QCD jet radiation on $\meff$. In order to assess this uncertainty, alternative samples were produced with reduced initial parton multiplicities (\alpgen~processes with 0--5 partons rather than 0--6 partons for $W/Z$+jets production, and 0--3 instead of 0--5 for top quark pair production). 

PDF uncertainties are also taken into account. An envelope of cross section predictions is defined using the 68\% confidence level (CL) ranges of the \cteq~\cite{Nadolsky:2008zw} (including the $\alpha_s$ uncertainty) and \mstw~\cite{Martin:2009iq} PDF sets, together with independent variations of the factorisation and renormalisation scales by factors of two or one half. The nominal cross section value is taken to be the midpoint of the envelope and the uncertainty assigned is half the full width of the envelope, closely following the PDF4LHC recommendations~\cite{Botje:2011sn}.

Additional uncertainties arising from photon and lepton reconstruction efficiency, energy scale and resolution in CR1a, CR1b, CR3 and CR4, $b$-tag/veto efficiency (CR3 and CR4) and photon acceptance and cosmic ray backgrounds (CR1a) are also considered. Other sources, including the limited number of MC simulation events as well as
additional systematic uncertainties related to the response function, are included.

Systematic uncertainties on the expected SUSY signal are estimated through variation
of the factorisation and renormalisation scales between half
and twice their default values and by considering the PDF uncertainties.
Uncertainties are calculated for individual
production processes (e.g.~$\squark\squark$, $\gluino\gluino$, etc.). 

Initial state radiation (ISR) can significantly affect the signal visibility for SUSY models with small mass splittings. Systematic uncertainties arising from the treatment of ISR are studied by varying the assumed value of $\alpha_s$ and the {\tt MadGraph}/{\pythia} matching parameters. The uncertainties are found to be negligible for large sparticle masses ($m > 300$~GeV) and mass splittings ($\Delta m > 300$~GeV), and to rise linearly with decreasing mass and decreasing mass splitting to $\sim$30\% for  $\Delta m = 0$ and $m > 300$~GeV, and to  $\sim$40\% for $m = 250$~GeV and $\Delta m =0$. Signal ISR uncertainties are assumed to be uncorrelated with the corresponding background ISR uncertainties, to ensure a conservative treatment.

\section{Results, Interpretation and Limits}
\label{sec:stats}

\begin{figure*}[t]
\begin{center}

\subfigure[]{\includegraphics[width=0.48\textwidth]{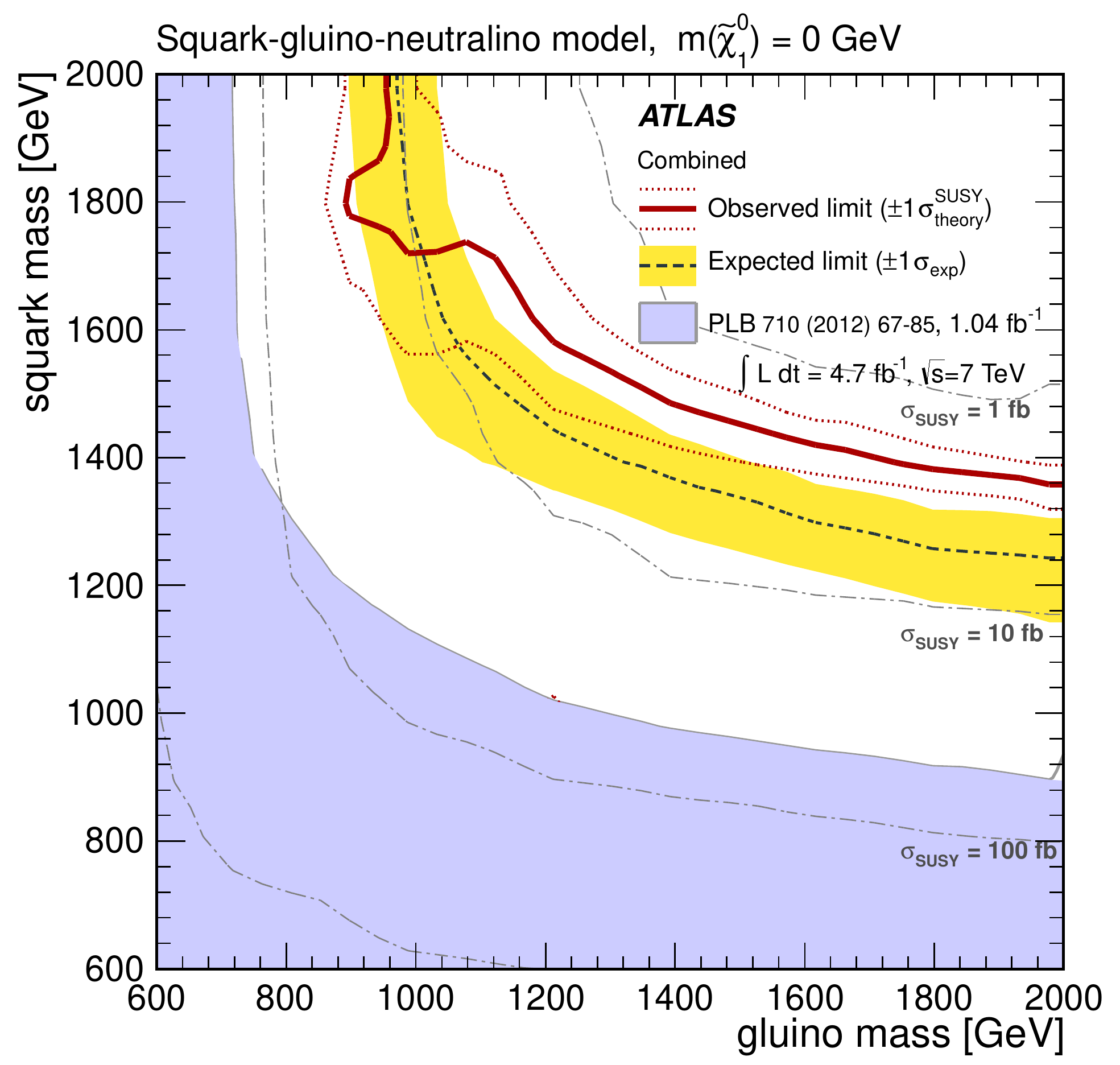}}
\subfigure[]{\includegraphics[width=0.48\textwidth]{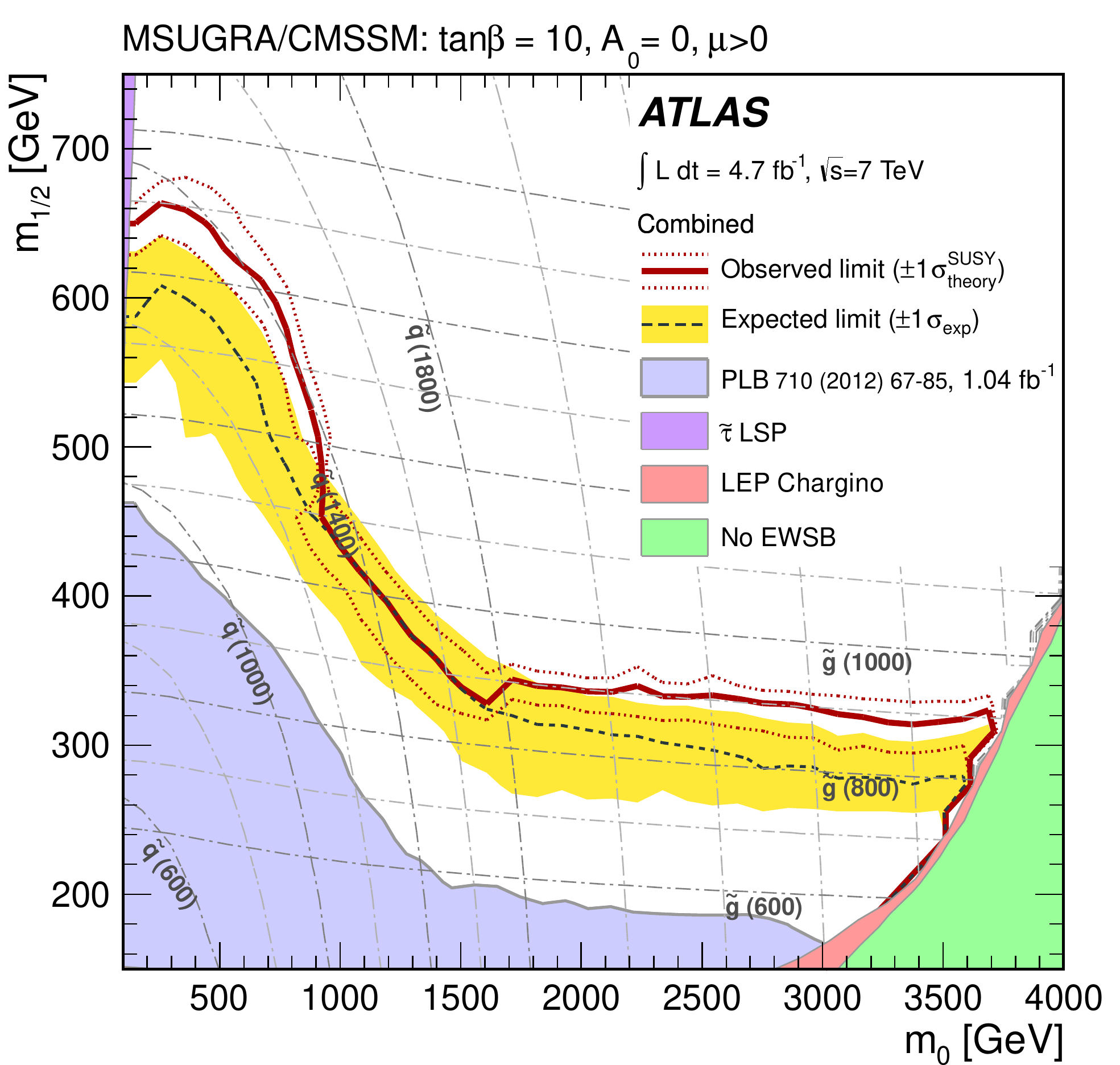}}

 \caption {The 95\% CL$_{\rm s}$ exclusion limits on (a) the $(m_{\gluino}, m_{\squark})$-plane in a simplified MSSM scenario with only strong production of gluinos and first- and second-generation squarks, with direct decays to jets and neutralinos; (b) the ($m_0, m_{1/2}$) plane of MSUGRA/CMSSM\ for $\tan\beta=10$, $A_0=0$ and $\mu>0$. Exclusion limits are obtained by using the signal region with the best expected sensitivity at each point. The black dashed lines show the expected limits, with the light (yellow) bands indicating the $1\sigma$ excursions due to experimental uncertainties.
Observed limits are indicated by medium (maroon) curves, where the solid contour represents the nominal limit, and the dotted lines are obtained by varying the cross section by the theoretical scale and PDF uncertainties.
Previous results from ATLAS~\cite{Aad:2011ib} are represented by the shaded region (blue) at bottom left in each case. The region excluded by chargino searches at LEP is taken from Ref.~\cite{Abdallah:681867}.
\label{fig:limitcombined}}
\end{center}
\end{figure*}

The numbers of events observed in the data and the numbers of SM events expected to enter the signal regions, determined using the simultaneous likelihood fits (see Sections~\ref{sec:strategy} and~\ref{sec:back}) to the SRs and CRs, are shown in Table~\ref{tab:SRev}. 
The use of transfer factors between the CRs and SRs allows systematic uncertainties and nuisance parameters to be dealt with in a coherent way, preserving any correlations, as described above. 
The free parameters are the background components in each SR, and these are constrained by the CR event counts and the TFs, within their uncertainties. The dominant irreducible background, from $Z$+jets events, is constrained by both CR1a and CR1b, with CR1a providing the largest statistical weight. The resulting scaled predictions for the background components are shown in Table~\ref{tab:SRev}.
Good agreement is observed between the data and the SM predictions, with no significant excesses found. The fitted predictions for the various background components agree well with the expectations from MC simulation before the fits, once theoretical uncertainties are accounted for.

Data from all the channels are used to set limits on SUSY models, taking the SR with the best expected sensitivity at each point in parameter space. A profile log-likelihood ratio test in combination with the CL$_{\rm s}$ prescription~\cite{clsread} is used to derive 95\% CL exclusion regions.
An interpretation of the results is presented in
Fig.~\ref{fig:limitcombined}{}(a) as a 95\% CL exclusion
region in the $(m_{\gluino},m_{\squark})$-plane for a set of simplified SUSY
models with $m_{\ninoone}=0$.  In these models the gluino mass and the masses of the
squarks of the first two generations are set to the values shown in the figure, up to maximum squark and gluino masses of 2~TeV. All other supersymmetric particles,
including the squarks of the third generation, are decoupled. 
The results are also interpreted in the
$\tan\beta=10$, $A_0=0$, $\mu>0$ slice of MSUGRA/CMSSM~models~\cite{msugrapars} 
in Figure~\ref{fig:limitcombined}(b).
In these models, \isasusy{} from
\isajet{} \cite{Paige:2003mg} {v7.80} is used to calculate the
decay tables, and to guarantee consistent electroweak symmetry
breaking.

In the simplified model with light neutralinos, with the assumption that the colored sparticles are directly produced and
decay directly to jets and $\met$, the limit on the gluino mass is
approximately 860~GeV, and that on the squark mass is 1320~GeV. Squarks and gluinos with equal masses below 1410~GeV are excluded. These values are derived from the lower edge of the $1\sigma$ observed limit band, to take account of the theoretical uncertainties on the SUSY cross sections in a conservative fashion. In the MSUGRA/CMSSM case, the limit on $m_{1/2}$ reaches 300~GeV at high $m_0$ and 640~GeV for
low values of $m_0$. The inclusion of signal selections sensitive to larger jet multiplicities
has improved significantly the ATLAS reach at large $m_0$. When their masses are assumed to be equal, squarks and gluinos with masses below 1360~GeV are excluded.

\begin{figure*}
\begin{center}

\subfigure[]{\includegraphics[width=0.48\textwidth]{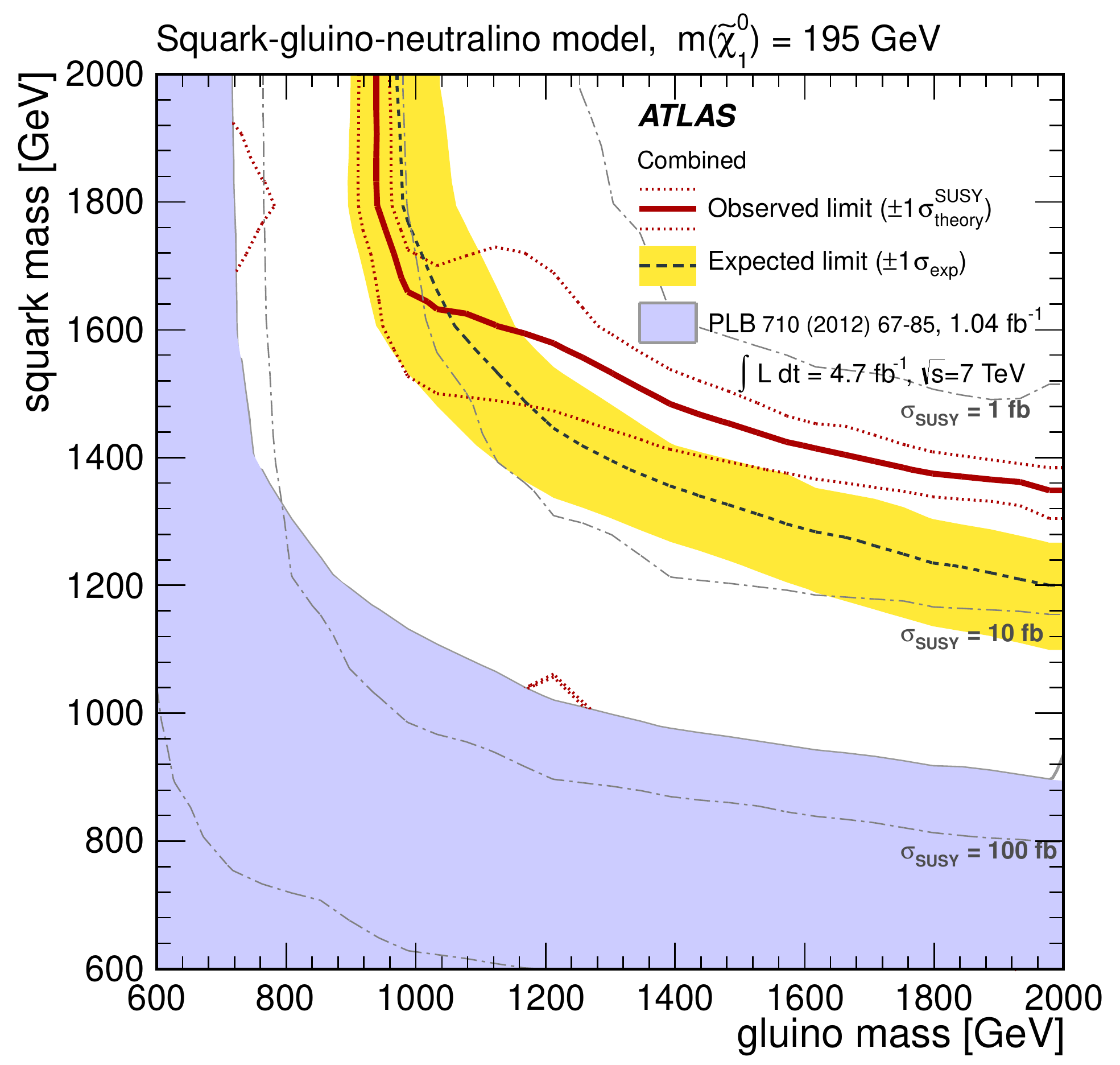}}
\subfigure[]{\includegraphics[width=0.48\textwidth]{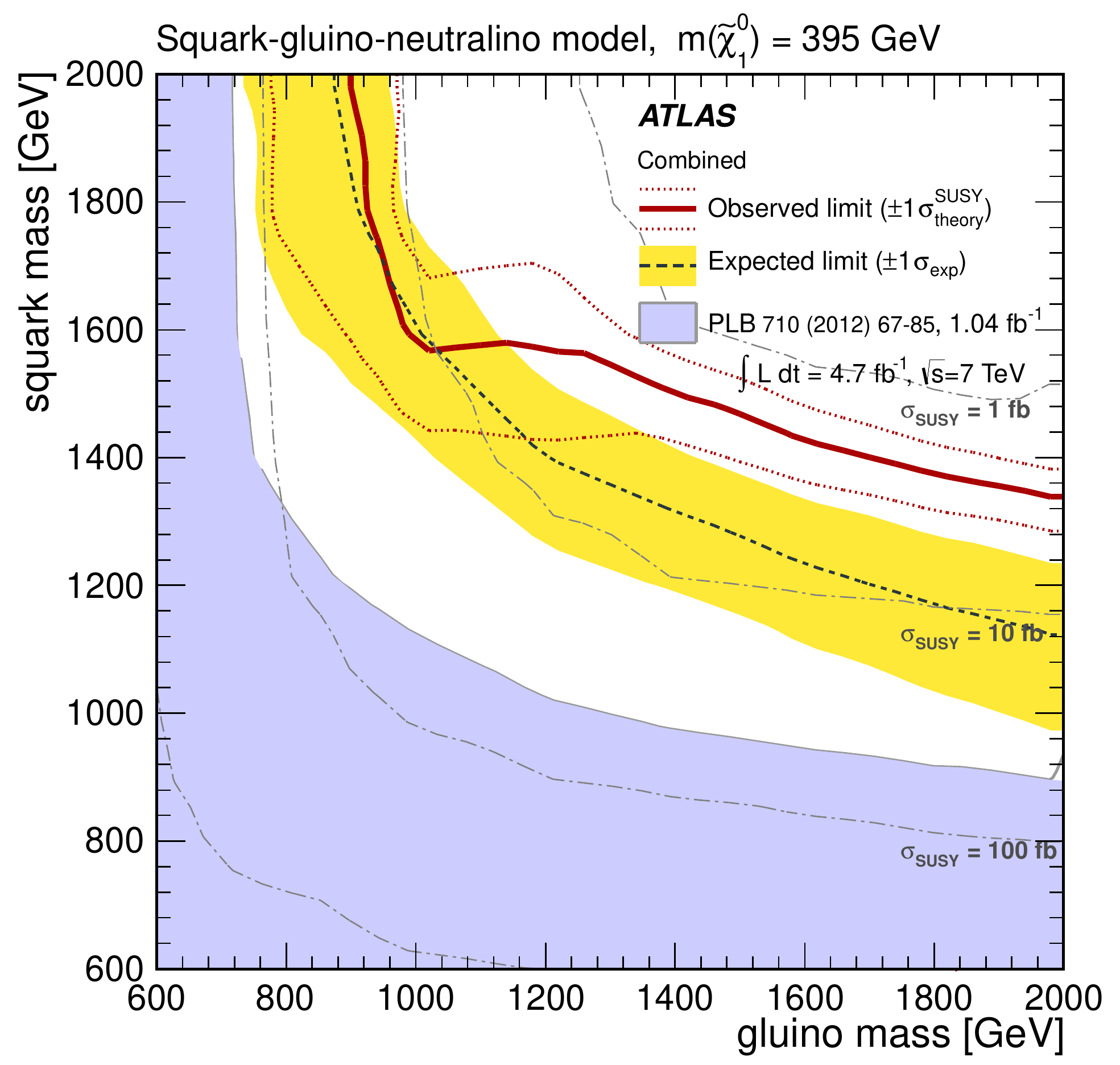}}

 \caption{ The 95\% CL$_{\rm s}$ exclusion limits on the $(m_{\gluino}, m_{\squark})$-plane in MSSM models with non-zero neutralino masses.
Combined observed exclusion limits are based on the best expected CL$_{\rm s}$ per grid point as for Fig.~\ref{fig:limitcombined}(a), but with an LSP mass of (a) 195~GeV and (b) 395~GeV. Curves are as defined in Fig.~\ref{fig:limitcombined}(a). The letters overlaid on the plot show the SR that contributes the best sensitivity at each point.
Previous results from ATLAS~\cite{Aad:2011ib} are represented by the shaded region (blue) at bottom left in each case.
\label{fig:limitmlspnezerocombined_labelled}}
\end{center}
\end{figure*}

In Figures~\ref{fig:limitmlspnezerocombined_labelled}(a) and~\ref{fig:limitmlspnezerocombined_labelled}(b) the limits from Fig.~\ref{fig:limitcombined}(a) are displayed again, but with the LSP mass set to 195~GeV and 395~GeV respectively. For both values, only minor differences are seen in the limit curve, showing that the analysis retains sensitivity for a range of LSP masses. The signal region with the greatest reach is displayed at each point in the plane, showing that the tight, medium and loose selections all contribute to the final result.

\begin{figure*}
\begin{center}
\subfigure[]{\includegraphics[width=0.48\textwidth]{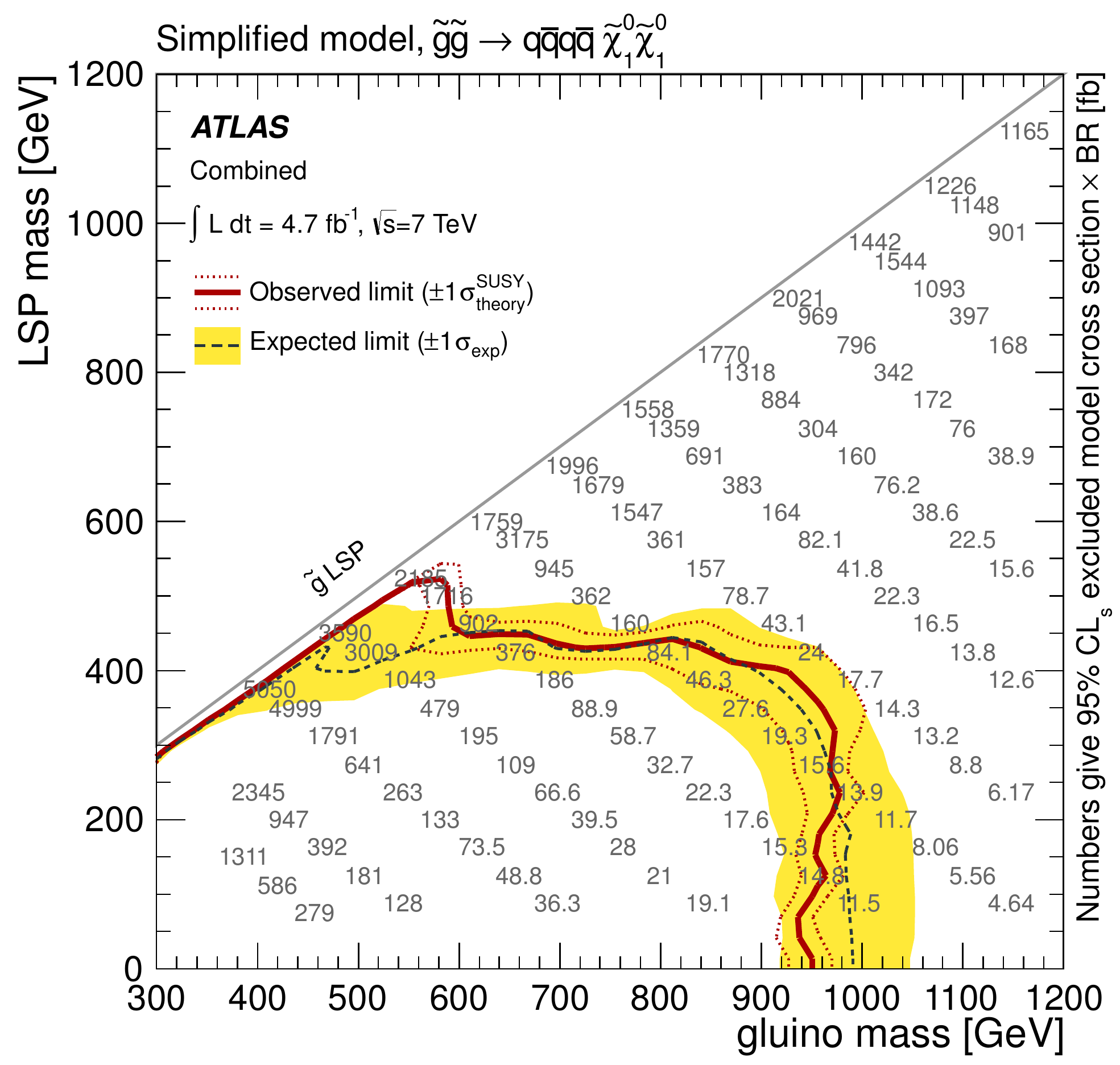}}
\subfigure[]{\includegraphics[width=0.48\textwidth]{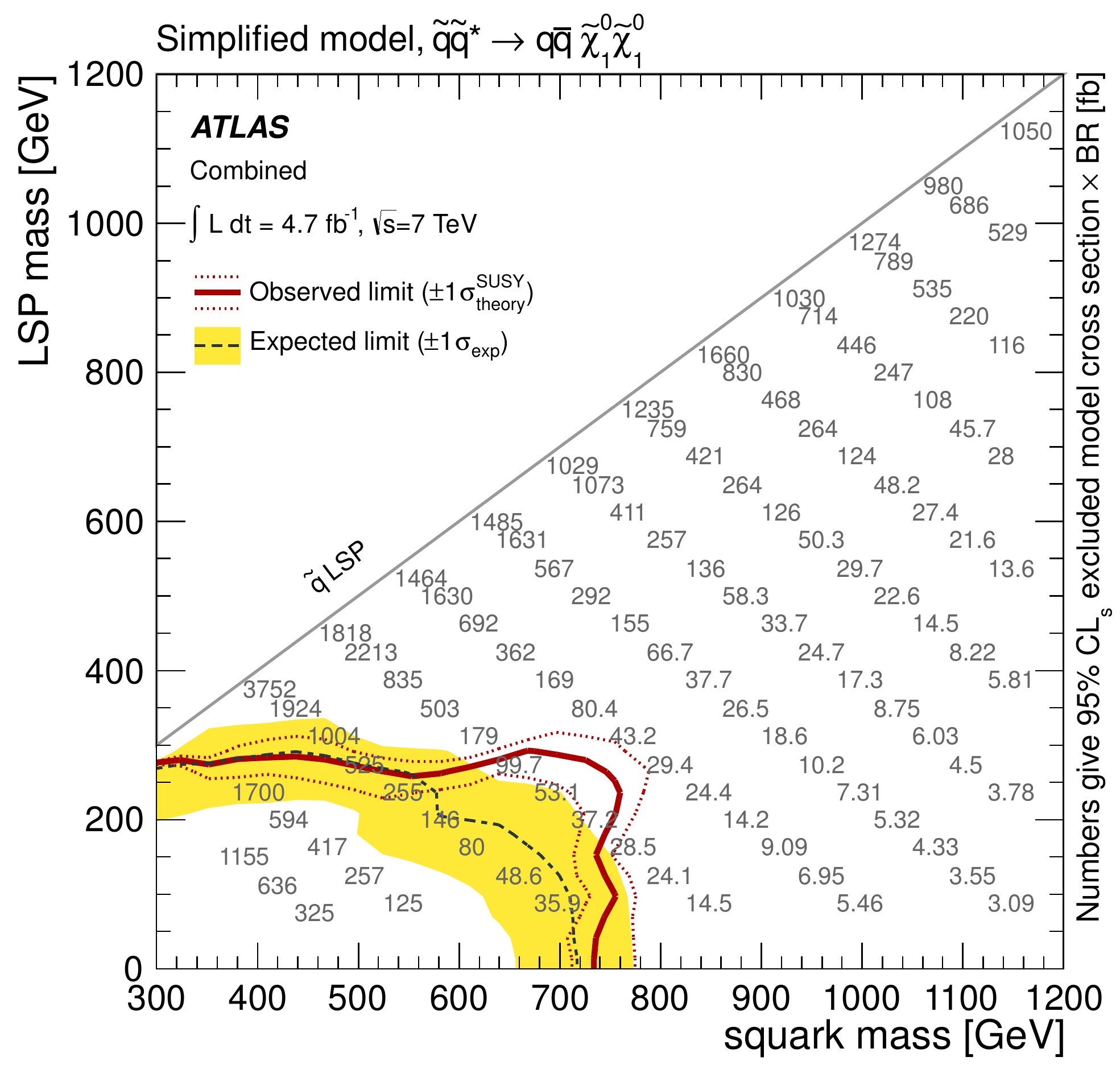}}
 \caption{ The 95\% CL$_{\rm s}$ exclusion limits on simplified models assuming direct production of (a) gluino pairs with decoupled squarks or (b) squark pairs with decoupled gluinos, each decaying to two jets, or one jet, respectively, and a neutralino LSP. 95\% Exclusion limits are obtained by using the signal region with the best expected sensitivity at each point.  
The black dashed lines show the expected limits, with the light (yellow) bands indicating the $1\sigma$ excursions due to experimental uncertainties.
Observed limits are indicated by medium (maroon) curves, where the solid contour represents the nominal limit, and the dotted lines are obtained by varying the cross section by the theoretical scale and PDF uncertainties.
The 95\% CL$_{\rm s}$ upper limit on the cross section times branching ratio (in fb) is printed for each model point. 
\label{fig:directLimit}}
\end{center}
\end{figure*}

In Figure~\ref{fig:directLimit} limits are shown for two cases in which only pair production of (a) gluinos or (b) squarks is kinematically possible, with all other superpartners, except for the neutralino LSP, decoupled. This forces each squark or gluino to decay directly to jets and an LSP, as in the simplified MSSM scenario. Cross sections are evaluated assuming decoupled squarks or gluinos in cases (a) and (b), respectively.

\begin{figure*}
\begin{center}
\subfigure[]{\includegraphics[width=0.48\textwidth]{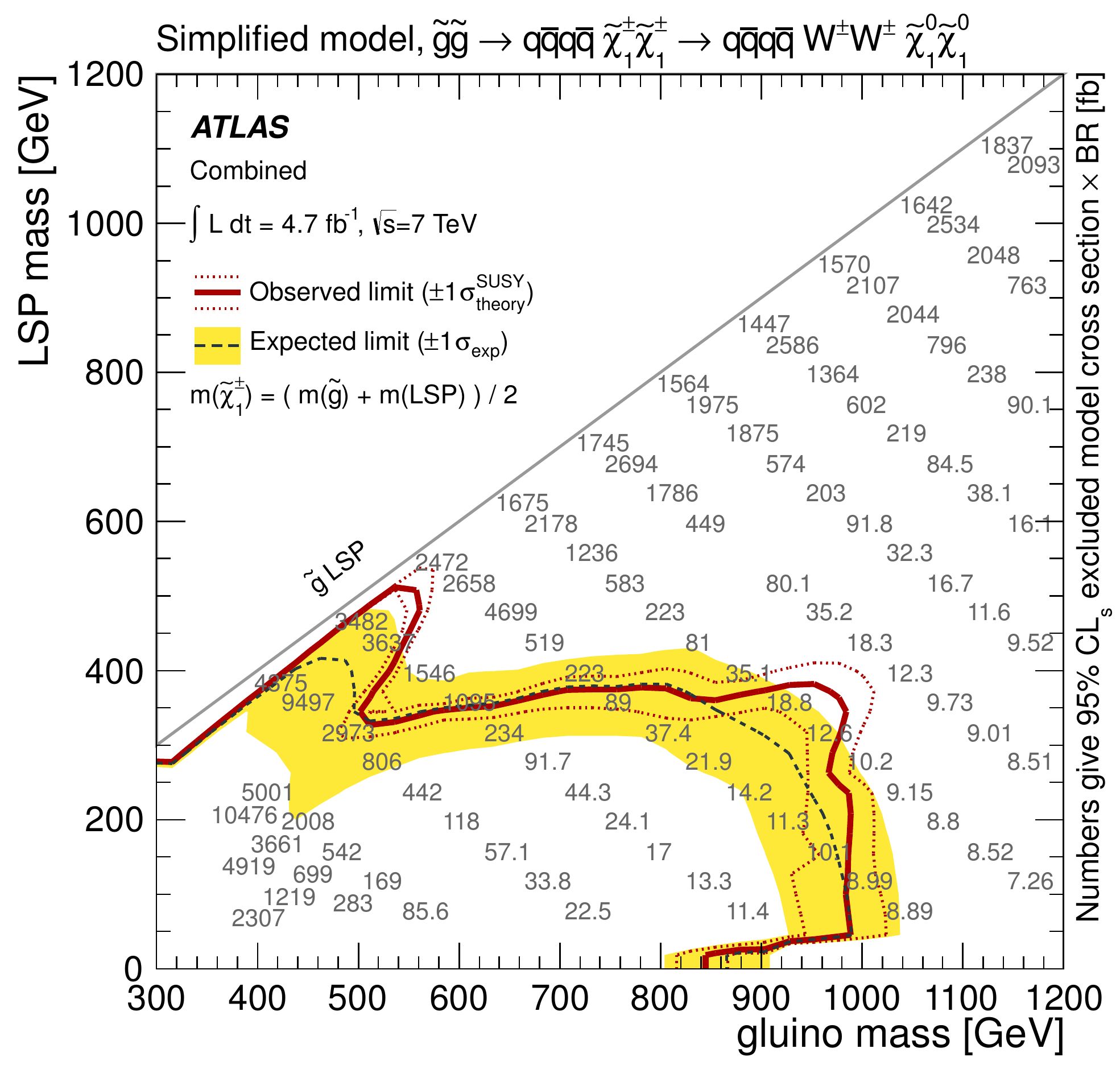}}
\subfigure[]{\includegraphics[width=0.48\textwidth]{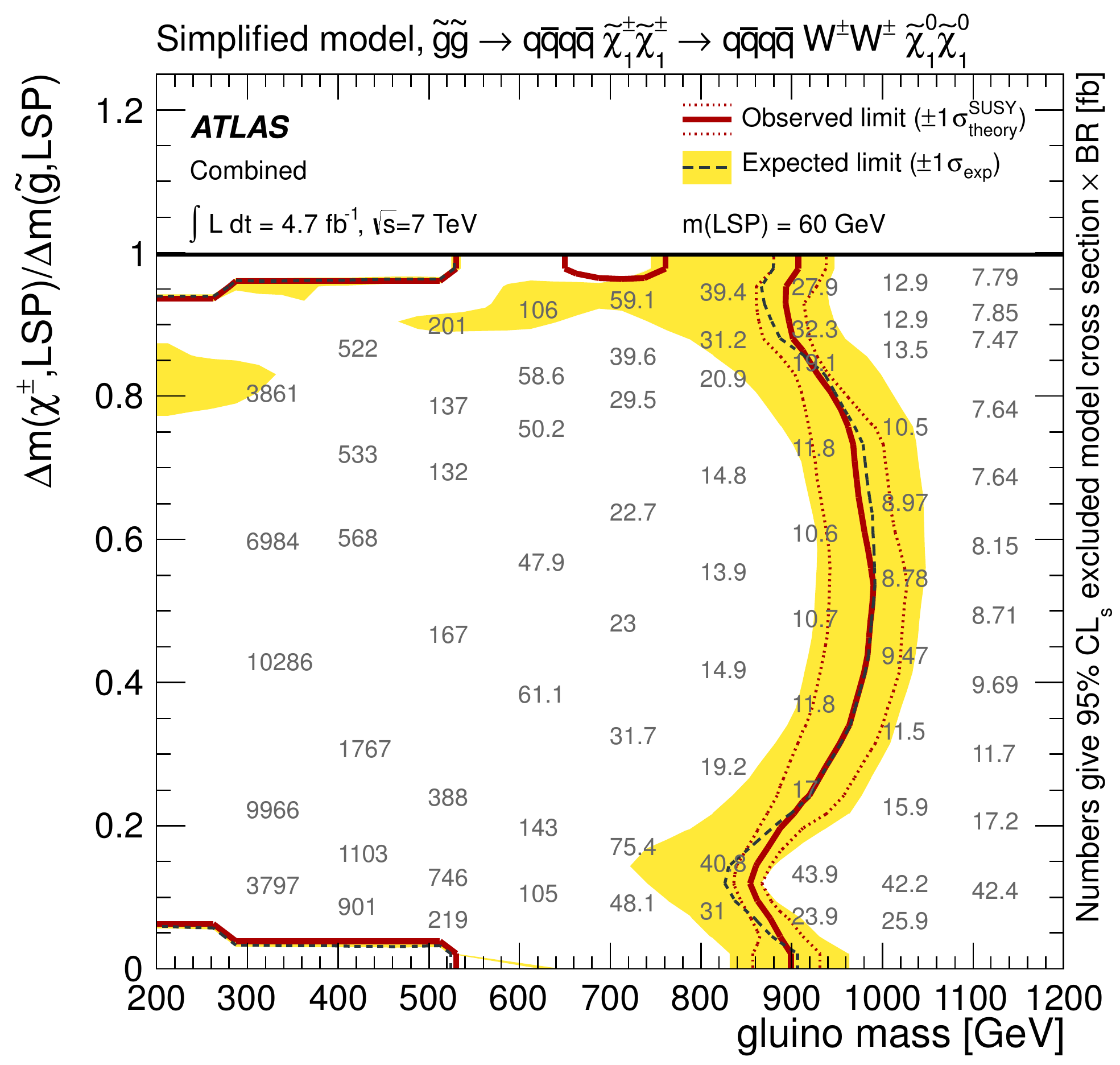}}
 \caption{ Combined 95\% CL$_{\rm s}$ exclusion limits on simplified models assuming direct production of gluino pairs, each decaying via an intermediate chargino to two jets, a $W$ boson and a neutralino LSP. The chargino mass is fixed halfway in between the gluino and LSP masses in figure (a). The neutralino mass is fixed at 60~GeV in figure (b), where the $y$-axis shows the ratio of the chargino-LSP mass-splitting to the gluino-LSP mass-splitting. The black dashed lines show the expected limits, with the light (yellow) bands indicating the $1\sigma$ excursions due to experimental uncertainties.
Observed limits are indicated by medium (maroon) curves, where the solid contour represents the nominal limit, and the dotted lines are obtained by varying the cross section by the theoretical scale and PDF uncertainties. The 95\% CL$_{\rm s}$ upper limit on the cross section times branching ratio (in fb) is printed for each model point. 
\label{fig:limitgluinocombined}}
\end{center}
\end{figure*}

\begin{figure*}
\begin{center}
\subfigure[]{\includegraphics[width=0.48\textwidth]{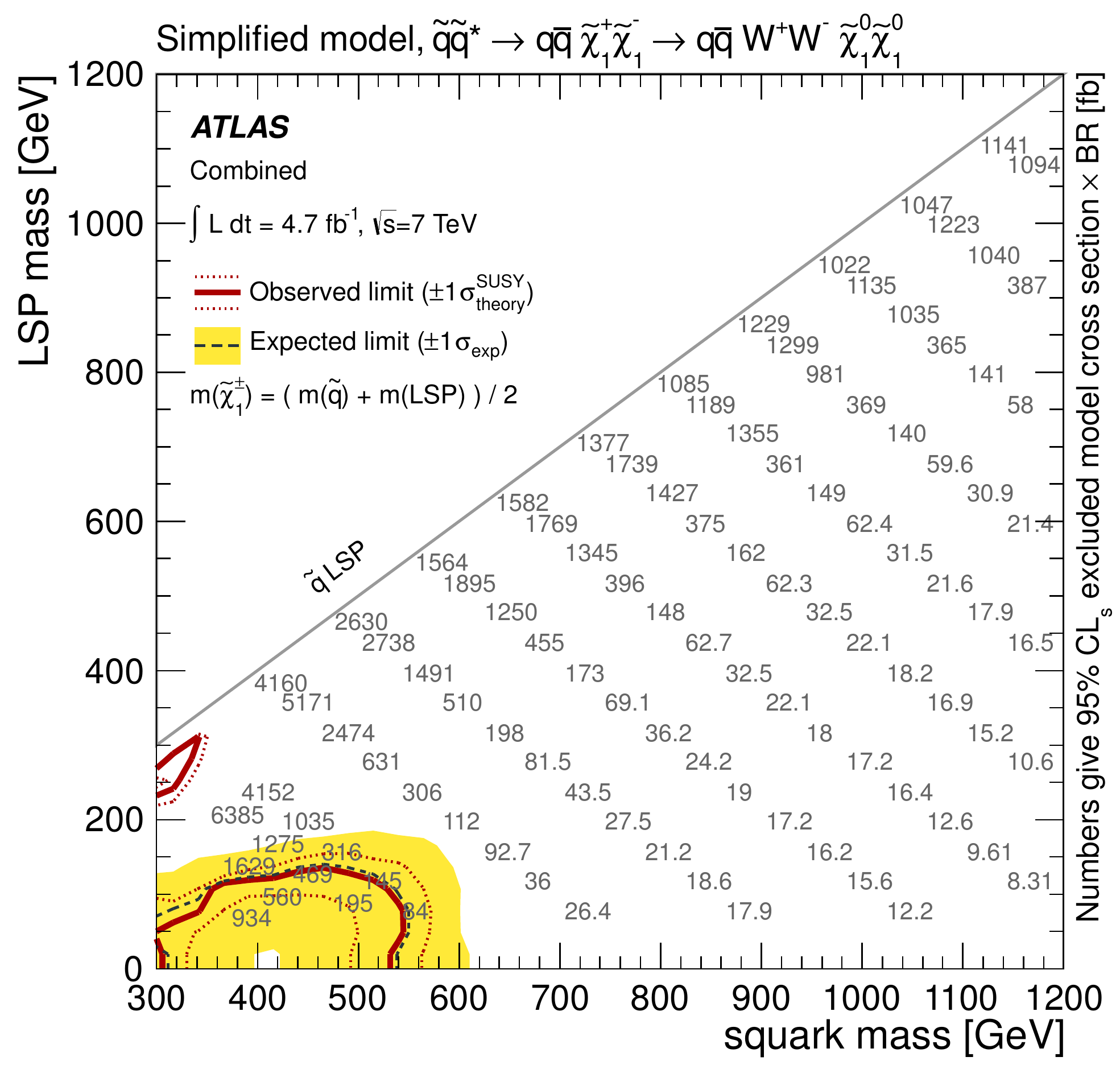}}
\subfigure[]{\includegraphics[width=0.48\textwidth]{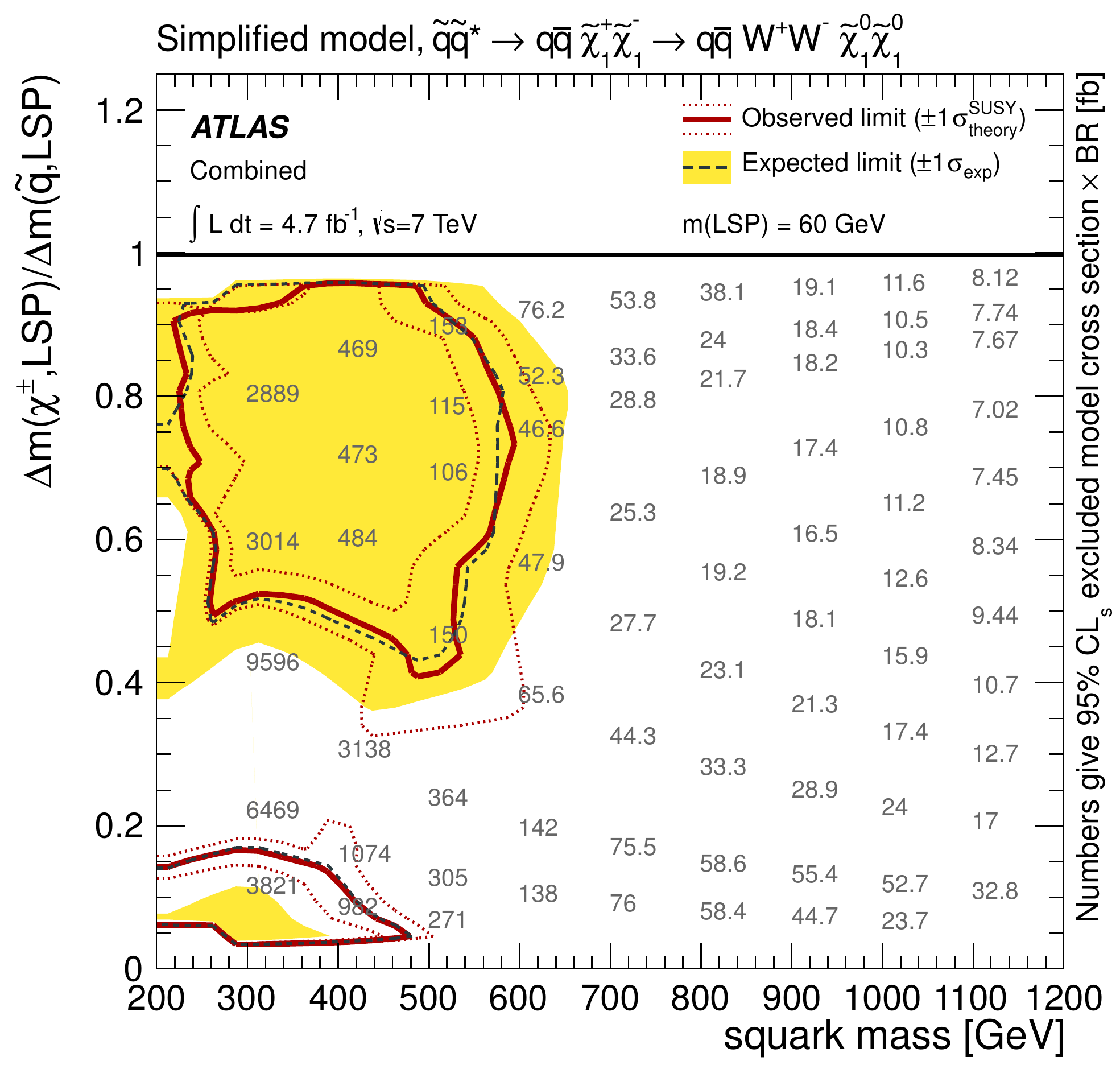}}
 \caption{ Combined 95\% CL$_{\rm s}$ exclusion limits on simplified models assuming direct production of left-handed squark-antisquark pairs, each decaying via an intermediate chargino to two jets, a $W$ boson and a neutralino LSP. The chargino mass is fixed halfway in between the squark and LSP masses in figure (a). In figure (b) the neutralino mass is fixed at 60~GeV; the $y$-axis shows the ratio of the chargino-LSP mass-splitting to the squark-LSP mass-splitting. The black dashed lines show the expected limits, with the light (yellow) bands indicating the $1\sigma$ excursions due to experimental uncertainties.
Observed limits are indicated by medium (maroon) curves, where the solid contour represents the nominal limit, and the dotted lines are obtained by varying the cross section by the theoretical scale and PDF uncertainties.  The 95\% CL$_{\rm s}$ upper limit on the cross section times branching ratio (in fb) is printed for each model point. 
\label{fig:limitsquarkcombined}}
\end{center}
\end{figure*}

Similar models with only squark or gluino pair-production are shown in Figs.~\ref{fig:limitgluinocombined} and \ref{fig:limitsquarkcombined}. However, in these variants, the sparticle content is augmented by an additional intermediate chargino with mass between the strongly-interacting sparticle and the LSP. This allows for production of additional jets or leptons and enriches the phenomenology. In the squark pair-production case, only left-handed squarks of the first and second generations are considered in order to enhance the branching ratios of decay chains incorporating an intermediate chargino. The cross sections have been reduced by 50\% to take this situation properly into account. Two different parameterizations of the masses are shown. Figures~\ref{fig:limitgluinocombined}(a) and \ref{fig:limitsquarkcombined}(a) vary the squark/gluino mass and the LSP mass, keeping the chargino mass exactly midway between those two. In Figures~\ref{fig:limitgluinocombined}(b) and \ref{fig:limitsquarkcombined}(b), the LSP mass is instead held fixed, with the ratio of the chargino-LSP mass splitting to the squark/gluino-LSP mass splitting defining the $y$-axis. When either mass splitting is large sensitivity to the model is enhanced by kinematics.

\begin{figure*}
\begin{center}
\subfigure[]{\includegraphics[width=0.48\textwidth]{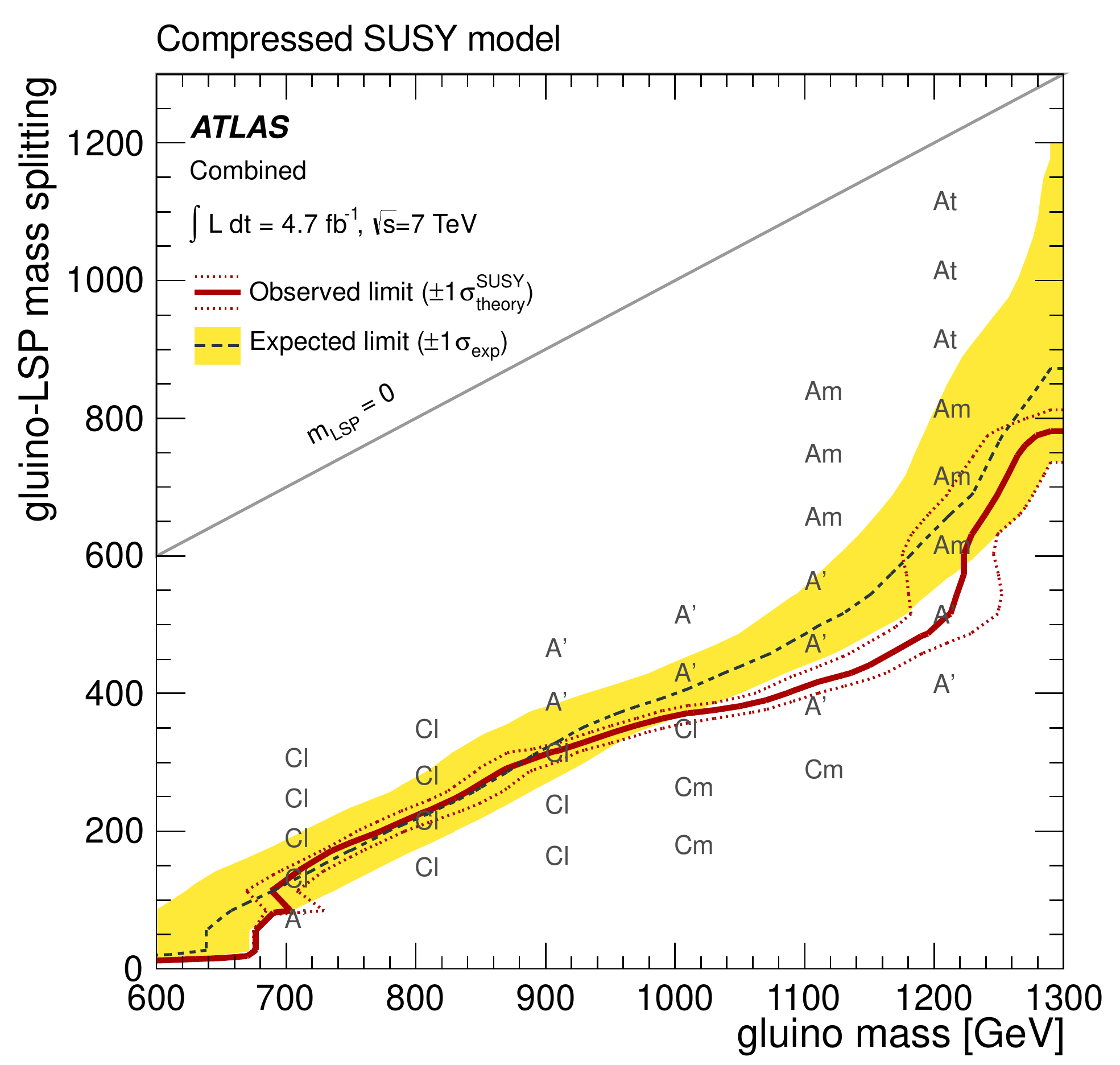}}
\subfigure[]{\includegraphics[width=0.48\textwidth]{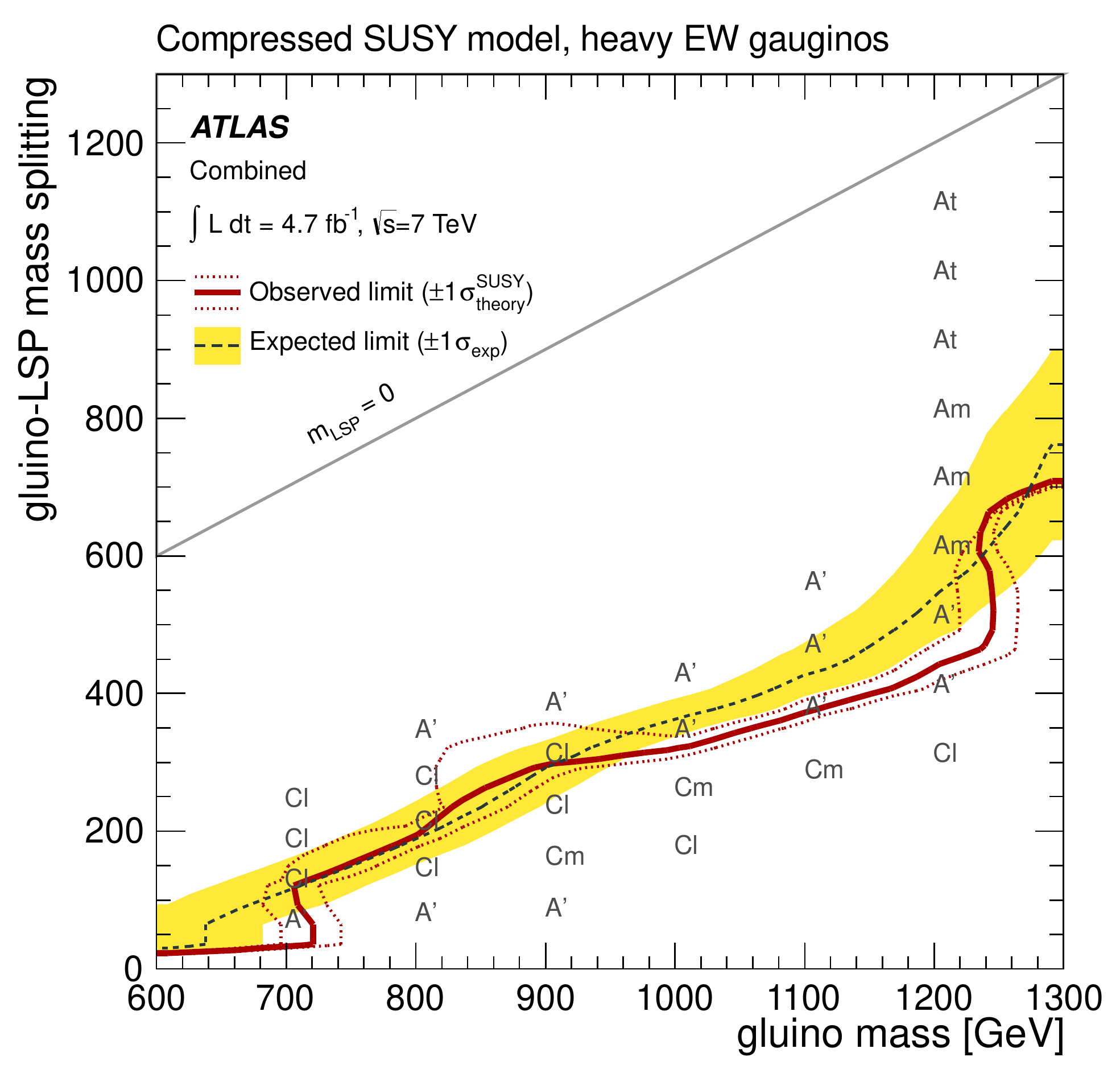}}
\subfigure[]{\includegraphics[width=0.48\textwidth]{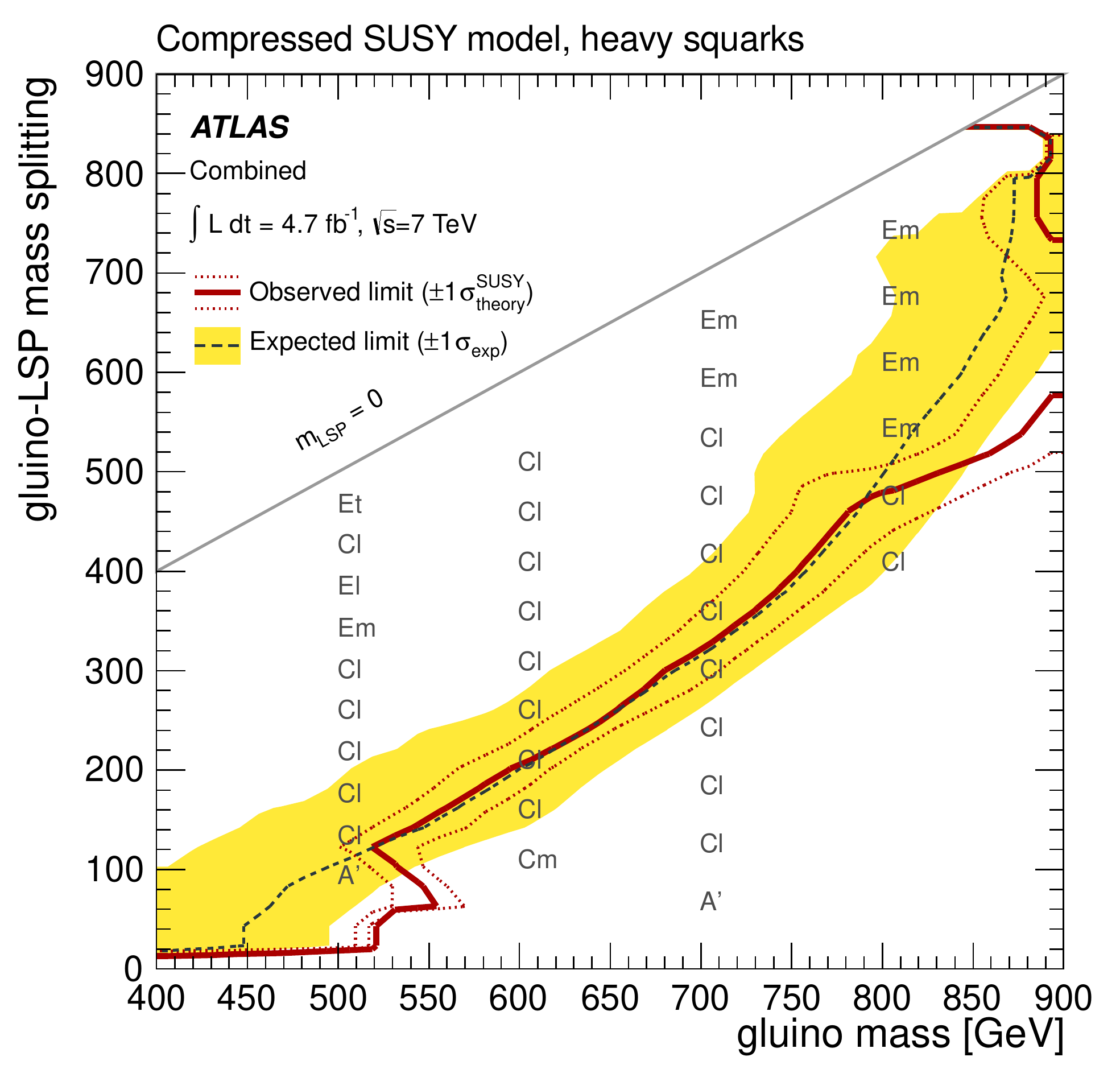}}

 \caption{
Combined 95\% CL$_{\rm s}$ exclusion limits for the compressed SUSY models discussed in the text. In figure (a) all squarks, electroweak gauginos and the gluino are kinematically accessible. In figure (b) neutralinos (apart from the LSP) and charginos are decoupled. In figure (c) squarks are decoupled.
The black dashed lines show the expected limits, with the light (yellow) bands indicating the $1\sigma$ excursions due to experimental uncertainties.
Observed limits are indicated by medium (maroon) curves, where the solid contour represents the nominal limit, and the dotted lines are obtained by varying the cross section by the theoretical scale and PDF uncertainties.
The letters overlaid on the plot show the SR that contributes the best sensitivity at each point.
\label{fig:compSUSYLimit}}
\end{center}
\end{figure*}

The `compressed SUSY'  models suggested in Refs.~\cite{TomSteve1, TomSteve2} are also considered. In these models, the basic sparticle content and spectrum are similar to that in the CMSSM, but the sizes of all mass-splittings are controlled by a compression factor. The squark mass is set to 96\% of the gluino mass. For presentation purposes, the limits are plotted against the gluino mass and the largest mass-splitting, i.e. that between gluino and LSP.
Exclusion plots are shown in Fig.~\ref{fig:compSUSYLimit} for three classes of model: one in which all sparticle content is present, a second in which all the neutralinos and charginos apart from the LSP are taken to be sufficiently heavy to decouple, and a third in which the squarks instead are decoupled. 

\section{Summary}
\label{sec:conc}

This paper reports a search for supersymmetry in final states containing
high-\pT{} jets, missing transverse momentum and no electrons with $\ourpt >20$~GeV or muons with $\ourpt >10$~GeV. Data recorded by the ATLAS experiment at the LHC at $\sqrt{s}=7$~TeV, corresponding to an integrated luminosity of \ourintlumi{}  have been used.
Good agreement is seen between the numbers of events observed in the signal regions and the numbers of events expected from SM sources. 
The exclusion limits placed on non-SM cross sections impose new constraints on scenarios with novel physics.

The results are interpreted in both a simplified model containing
only squarks of the first two generations, a gluino octet and a
massless neutralino, as well as in MSUGRA/CMSSM models with $\tan\beta=10$, $A_0=0$
and $\mu>0$. 
In the simplified model, gluino and squark masses below 860~GeV and 1320~GeV respectively
are excluded at the $95\%$ confidence level for squark or gluino masses below 2~TeV.
When assuming their masses to be equal, squarks and gluinos with masses below 1410~GeV are excluded. 
In the MSUGRA/CMSSM case, the limit on $m_{1/2}$ reaches 300~GeV at high $m_0$ and 640~GeV for
low values of $m_0$. 
Squarks and gluinos with equal masses below 1360~GeV are excluded in this scenario. 
These results are shown to be relatively insensitive to the assumption of a light LSP, up to LSP masses of about 400~GeV. 
Limits are also placed in the parameter space of a SUSY model with a compressed mass spectrum.

\section{Acknowledgements}

We thank CERN for the very successful operation of the LHC, as well as the
support staff from our institutions without whom ATLAS could not be
operated efficiently.

We acknowledge the support of ANPCyT, Argentina; YerPhI, Armenia; ARC,
Australia; BMWF and FWF, Austria; ANAS, Azerbaijan; SSTC, Belarus; CNPq and FAPESP,
Brazil; NSERC, NRC and CFI, Canada; CERN; CONICYT, Chile; CAS, MOST and NSFC,
China; COLCIENCIAS, Colombia; MSMT CR, MPO CR and VSC CR, Czech Republic;
DNRF, DNSRC and Lundbeck Foundation, Denmark; EPLANET and ERC, European Union;
IN2P3-CNRS, CEA-DSM/IRFU, France; GNSF, Georgia; BMBF, DFG, HGF, MPG and AvH
Foundation, Germany; GSRT, Greece; ISF, MINERVA, GIF, DIP and Benoziyo Center,
Israel; INFN, Italy; MEXT and JSPS, Japan; CNRST, Morocco; FOM and NWO,
Netherlands; BRF and RCN, Norway; MNiSW, Poland; GRICES and FCT, Portugal; MERYS
(MECTS), Romania; MES of Russia and ROSATOM, Russian Federation; JINR; MSTD,
Serbia; MSSR, Slovakia; ARRS and MVZT, Slovenia; DST/NRF, South Africa;
MICINN, Spain; SRC and Wallenberg Foundation, Sweden; SER, SNSF and Cantons of
Bern and Geneva, Switzerland; NSC, Taiwan; TAEK, Turkey; STFC, the Royal
Society and Leverhulme Trust, United Kingdom; DOE and NSF, United States of
America.

The crucial computing support from all WLCG partners is acknowledged
gratefully, in particular from CERN and the ATLAS Tier-1 facilities at
TRIUMF (Canada), NDGF (Denmark, Norway, Sweden), CC-IN2P3 (France),
KIT/GridKA (Germany), INFN-CNAF (Italy), NL-T1 (Netherlands), PIC (Spain),
ASGC (Taiwan), RAL (UK) and BNL (USA) and in the Tier-2 facilities
worldwide.

\vspace{5mm}

\bibliographystyle{atlasnote}
\bibliography{CONF_0lepton2012}

\begin{thebibliography}{10}
\bibitem{atlascoordinates}
{ATLAS uses a right-handed coordinate system with its origin at the nominal
  interaction point in the center of the detector and the $z$-axis along the
  beam pipe. Cylindrical coordinates $(r,\phi)$ are used in the transverse
  plane, $\phi$ being the azimuthal angle around the beam pipe. The
  pseudorapidity $\eta$ is defined in terms of the polar angle $\theta$ by
  $\eta=-\ln\tan(\theta/2)$. The transverse momentum $\pt$ is defined in terms
  of the magnitude of the three momentum $|p|$ by $\pt = |p|\sin\theta$.}
\bibitem{emscale}
{The electromagnetic energy scale is the basic calorimeter signal scale for the
  ATLAS calorimeters. It has been established using test-beam measurements for
  electrons and muons to give the correct response for the energy deposited in
  electromagnetic showers, while it does not correct for the lower response of
  the calorimeter to hadrons.}
\bibitem{smartveto}
{For jets, the amount of transverse energy ($E_T$) lost in the dead region can
  be estimated from the energy depositions in the neighboring calorimeter
  cells. If this lost $E_T$ projected along the $\anactualptmissvecwithapinit$
  direction amounts to more than 10~GeV and constitutes more than 10\% of the
  \met, the event is rejected.}
\bibitem{msugrapars}
{Five parameters are needed to specify a particular MSUGRA/CMSSM\ model: the
  universal scalar mass, $m_0$, the universal gaugino mass $m_{1/2}$, the
  universal trilinear scalar coupling, $A_0$, the ratio of the vacuum
  expectation values of the two Higgs fields, $\tan\beta$, and the sign of the
  higgsino mass parameter, $\mu=\pm$.}
\end{thebibliography}

\clearpage

\onecolumngrid
% ATLAS Collaboration author list
% Data extracted on 30-Nov-2012 for paper reference SUSY-2011-20
%\documentclass[11pt]{article}
%\usepackage{a4wide}\begin{document}
\begin{flushleft}
{\Large The ATLAS Collaboration}

\bigskip

G.~Aad$^{\rm 47}$,
T.~Abajyan$^{\rm 20}$,
B.~Abbott$^{\rm 110}$,
J.~Abdallah$^{\rm 11}$,
S.~Abdel~Khalek$^{\rm 114}$,
A.A.~Abdelalim$^{\rm 48}$,
O.~Abdinov$^{\rm 10}$,
R.~Aben$^{\rm 104}$,
B.~Abi$^{\rm 111}$,
M.~Abolins$^{\rm 87}$
% ...
\bigskip
\\
% ...
$^{10}$ Institute of Physics, Azerbaijan Academy of Sciences, Baku, Azerbaijan\\
$^{11}$ Institut de F{\'\i}sica d'Altes Energies and Departament de F{\'\i}sica de la Universitat Aut{\`o}noma de Barcelona and ICREA, Barcelona, Spain\\
% ...
$^{20}$ Physikalisches Institut, University of Bonn, Bonn, Germany\\
% ...
$^{47}$ Fakult{\"a}t f{\"u}r Mathematik und Physik, Albert-Ludwigs-Universit{\"a}t, Freiburg, Germany\\
$^{48}$ Section de Physique, Universit{\'e} de Gen{\`e}ve, Geneva, Switzerland\\
% ...
$^{87}$ Department of Physics and Astronomy, Michigan State University, East Lansing MI, United States of America\\
% ...
$^{104}$ Nikhef National Institute for Subatomic Physics and University of Amsterdam, Amsterdam, Netherlands\\
% ...
$^{110}$ Homer L. Dodge Department of Physics and Astronomy, University of Oklahoma, Norman OK, United States of America\\
$^{111}$ Department of Physics, Oklahoma State University, Stillwater OK, United States of America\\
% ...
$^{114}$ LAL, Universit{\'e} Paris-Sud and CNRS/IN2P3, Orsay, France\\
% ...
\end{flushleft}

%\end{document}
% Created with /afs/cern.ch/user/a/atlaspo/svntools/Upload/xml2latex.py

\end{document}